\documentclass{aa}  

\usepackage{graphicx}
\usepackage{txfonts}
\usepackage{lipsum}
\usepackage{subcaption}         
\usepackage{lscape}             
\usepackage{placeins}           

\usepackage{booktabs}
\usepackage[table]{xcolor}
\usepackage{multirow}
\usepackage{makecell}
\usepackage{hyperref}
\hypersetup{
    colorlinks=true,
    linkcolor=blue,
    citecolor=blue,
    urlcolor=blue
}
\usepackage{caption}
\definecolor{lightred}{RGB}{255,230,230}
\definecolor{lightyellow}{RGB}{255,248,210}
\definecolor{lightgreen}{RGB}{225,245,225}
\definecolor{lightblue}{RGB}{225,240,255}
                                
\begin{document}

   \title{Dynamical evolution and surface accretion of DART impact ejecta in the (65803) Didymos system}


%
%
%

   \author{X. Fu\inst{1}\corrauth{Xiaoyu.Fu@liverpool.ac.uk}        
        \and N. Stronati\inst{1}\email{Nicolo.Stronati@liverpool.ac.uk}
        \and S. Soldini\inst{1}\email{Stefania.Soldini@liverpool.ac.uk}
        \and F. Ferrari\inst{2}\email{Fabio1.Ferrari@polimi.it}
        \and C. Giordano\inst{2}\email{Carmine.Giordano@polimi.it}
        \and P. Panicucci\inst{2}\email{Paolo.Panicucci@polimi.it}
        \and A. Rossi\inst{3}\email{A.Rossi@ifac.cnr.it}
        \and A. Campo Bagatin\inst{4}\email{acb@ua.es}
        \and M. Küppers\inst{5}\email{michael.kueppers@esa.int}
        }

   \institute{Department of Mechanical and Aerospace Engineering, University of Liverpool, Liverpool, L69 3BX, United Kingdom
   \and Department of Aerospace Science and Technology, Politecnico di Milano, Milan 20159, Italy
   \and IFAC-CNR, Via Madonna del Piano 10, 50019 Sesto Fiorentino, Italy
   \and Instituto de F{\'{\i}}sica Aplicada a las Ciencias y las Tecnolog{\'{\i}}as, Universidad de Alicante, 03690 Alicante, Spain
   \and European Space Agency (ESA), ESAC, Camino Bajo del Castillo s/n, Villanueva de la Cañada, 28692, Madrid, Spain}

   \date{Received July XX, 2026}

 
  \abstract
   {The DART spacecraft impacted Dimorphos, the small moonlet of Didymos binary system, on 26 September 2022. The impact ejected dust, fragments, and boulders into the near-binary environment. In November 2026, ESA's Hera mission is expected to arrive at the binary system to characterise both asteroids and investigate the post-impact consequences in detail.}
   {We aim to investigate the dynamical evolution of DART-generated impact ejecta and to quantify their surface accretion patterns within the Didymos binary system.} 
   {High-fidelity ejecta dynamics, including polyhedron asteroid gravity and solar radiation pressure with combined occultations, are constructed. The ejecta initial conditions are generated from the observation-constrained velocity--size distribution and ejecta-cone geometry. In total, 20 million trajectories are integrated to characterise the ejecta evolution and surface accretion.}
   {More than 93.5\% of DART-generated ejecta particles escape from the system within two years, while only approximately 0.002\% remain in the near-binary environment. The deposited layer on Dimorphos reaches the order of 1.5 mm at mid-to-low latitudes. On Didymos, the accreted layer is mostly thinner than 0.3 mm, but may reach 3--11.5 mm in a localised high-density region.}
   {Most DART-generated ejecta are removed from the binary system, while a small but dynamically meaningful subset remains near the system or accretes onto the asteroid surfaces. The surface accretion distribution is strongly controlled by the initial ejecta-cone geometry, especially the cone-axis direction.}


   \maketitle

\section{Introduction}
{\label{C1:Section1}}
On September 26, 2022, NASA's Double Asteroid Redirection Test (DART) spacecraft conducted a planned impact on Dimorphos, the small moonlet in the Didymos binary asteroid system \citep{Cheng2023Momentum}. This kinetic-impact experiment successfully shortened the mutual orbital period of the binary system by 33.0$\pm$1.0 minutes, demonstrating the viability of kinetic impact as a method for deflecting potentially hazardous asteroids \citep{Thomas2023Orbital}. Following the impact, the generated ejecta was first imaged by the LICIACube Cubesat, which revealed a complex conical plume containing dust, fragments, and larger boulders \citep{Dotto2024Dimorphos}. Subsequently, multiple space- and ground-based observations monitored the evolution of ejecta structures, including the expanding ejecta cloud and ejecta tails, over a period of several weeks \citep{Li2023Ejecta, Graykowski2023light}. Further observations by the Hubbule Space Telescope (HST) revealed the presence of dozens of boulders in the near-binary environment \citep{Jewitt2023Dimorphos}. Detailed investigation of the ejecta morphology and dynamical evolution provides important constraints on the physical properties of the asteroid material and the momentum-transfer efficiency of the impact \citep{Cheng2023Momentum}. Furthermore, understanding ejecta-evolution mechanism is essential for improving planetary-defence strategies and for supporting the safety and design of future asteroid missions \citep{Chabot2024achievement}.

The dynamics and evolution of DART ejecta have been extensively investigated from multiple perspectives. Prior to the DART impact, \cite{Yu2017Ejecta} developed a comprehensive dynamical model incorporating the relevant forces to simulate the resulting ejecta cloud. \cite{Rossi2022Dynamical} simulated the evolution of DART ejecta over various timescales and analysed their survivability in the near-binary environment. \cite{Ferrari2022Ejecta} also predicted the ejecta-cloud evolution by dividing the process into several dynamically connected phases. As observational data accumulated after the impact, studies of ejecta evolution expanded to address a wider range of physical and dynamical aspects. For instance, \cite{Langner2024Long} focused on the long-term evolution of boulders generated by the DART impact, estimating their lifetimes and orbit elements. Their simulations suggest that upcoming missions may have a low but non-negligible probability of detecting such boulders. The authors also examined the sesquinary impacts on the surface of Didymos and the potential distribution of small, shallow craters produced by these impacts \citep{Langner2025Secondary}. \cite{Trogolo2025Evolution} investigated an ejecta-evolution mechanism driven by the fast-spinning Didymos and proposed that this process could contribute to the ejecta detected in the binary system. \cite{Fu2025Orbital} studied the orbital capture of impact ejecta around the Didymos binary system, with results indicating that ejecta particles reaching the periodic orbits considered for Hera observations are unlikely to remain near these orbits or to pose a potential hazard to the spacecraft. \cite{Ferrari2025Morphology} reproduced the observed ejecta morphologies using ejecta-dynamics simulations and synthetic HST images. Their analysis also constrained the ejecta size-frequency-distribution slope and total mass from HST observations.

Recent studies have substantially advanced the understanding of DART ejecta evolution, yet crucial modelling uncertainties and inaccuracies remain. Most previous simulations have relied on simplified representations of the binary asteroids, including approximate ellipsoidal shape models and reduced perturbation environments. While these assumptions are often appropriate for first-order interpretation and computational efficiency, they may affect both qualitative trends and quantitative estimates of ejecta fate, residence time, and surface accretion patterns. In addition, the partially constrained ejecta-cone geometries and the inaccurate impact-site location adopted in previous research can further influence the simulated dynamical evolution, particularly when assessing surface-impact probabilities and accretion-density distributions.

Motivated by the latest observation-based constraints on the DART ejecta cone \citep{Dotto2024Dimorphos, Hirabayashi2025elliptical}, this work incorporates updated ejecta-cone geometries, size--speed and size--frequency distributions, mission-derived ephemerides, and high-fidelity acceleration modelling. Large-scale particle simulations are then performed to obtain statistically robust trends in ejecta evolution and to quantify the fate distribution and surface accretion patterns of DART-generated ejecta. In this manner, the study aims to bridge observationally constrained ejecta properties with long-term dynamical modelling of the Didymos binary environment. 

This work is also timely in the context of ESA's Hera mission, which is expected to arrive at the Didymos system in November 2026 \citep{Michel2022ESA}. Hera will conduct in-situ characterisation of Didymos and Dimorphos and will provide critical post-impact measurements of the DART impact outcome, potentially including evidence of ejecta redistribution and surface accretion within the binary system. The high-fidelity ejecta-evolution simulations presented in this work may therefore provide useful dynamical context for interpreting Hera observations and for assessing how DART-generated material is transported, retained, or accreted in the Didymos system.

This paper is organised as follows. Sect.~\ref{C2:Section2} presents the dynamical model used to propagate DART ejecta particles after their departure from the surface of Dimorphos. The methodology for generating the initial conditions of the ejecta particles is then described in Sect.~\ref{C3:Section3}. Sect.~\ref{C4:Section4} provides a detailed statistical analysis of the DART ejecta evolution simulations. Based on the simulation results, two types of surface accretion-density maps are generated and analysed in Sect.~\ref{C5:Section5}. Finally, the conclusions are drawn in Sect.~\ref{C6:Section6}.

\section{Ejecta dynamics}
{\label{C2:Section2}}
This section describes the dynamical model used to propagate ejecta particles after their departure from the surface of Dimorphos. A binary-barycentric ecliptic J2000 (ECLIPJ2000) frame is adopted to describe the motion of an ejecta particle. The governing accelerations are then detailed separately, including the gravitational accelerations due to the polyhedral representations of Didymos and Dimorphos, third-body perturbation from the Sun and major planets, and solar radiation pressure (SRP) acceleration with combined occultations.

\subsection{Ejecta dynamics in the ECLIPJ2000 frame}
{\label{C21:SubSection21}}

In this study, the motion of an ejecta particle is modelled in an ECLIPJ2000 frame centred at the barycentre of the Didymos binary system. In this quasi-inertial frame, the $x$-axis is aligned with the J2000 mean vernal-equinox direction, i.e., the intersection of the Earth's mean equatorial plane and mean ecliptic plane at the J2000 epoch; the $z$-axis is normal to the mean ecliptic plane and points towards the ecliptic north pole; and the $y$-axis completes the right-handed coordinate system \citep{Vallado2001Fundamentals}. Under the definition above, the binary-barycentric ECLIPJ2000 frame differs from the solar system barycentre (SSB) inertial frame only by an accelerating translation of the origin, with no relative rotation between the two frames. Since the characteristic propagation time considered in this study is shorter than the heliocentric orbital period of the Didymos system, the resulting non-inertial effects are assumed to be negligible, and this barycentric ECLIPJ2000 frame is treated as quasi-inertial for the purpose of modelling the ejecta dynamics. 

To formulate ejecta dynamics in the designated ECLIPJ2000 frame, an ejecta particle is assumed to be homogenous and spherical, and the mutual gravitational interactions among particles are neglected. Consequently, the motion of an ejecta particle is driven mainly by the gravitational attraction of Didymos and Dimorphos, third-body perturbations from the Sun and major planets, and SRP. Accordingly, the acceleration of an ejecta particle is given as follows:
\begin{equation}
    \ddot{\boldsymbol{r}}_{\text{EJ2000}} =  \Sigma_{i=1}^{2} \boldsymbol{a}^{D_i}_{\text{EJ2000}} + \Sigma_{j=1}^{n} \boldsymbol{a}^{P_j}_{\text{3rd}} + \boldsymbol{a}_{\text{SRP}}
\label{eq:EOM_EJ2000}
\end{equation}
where $\boldsymbol{r}_{\text{EJ2000}}=[x,y,z]^T$ denotes the ejecta's position vector in the ECLIPJ2000 frame.\footnote{For notational convenience, the conventional abbreviation ``ECLIPJ2000'' is shortened to ``EJ2000'' when used as a suffix in vector and acceleration terms. This notation is adopted throughout the remainder of this manuscript.} The terms $\boldsymbol{a}^{D_i}_{\text{EJ2000}}\;(i=1,2)$ denote the gravitational accelerations due to Didymos ($D_1$) and Dimorphos ($D_2$), respectively. $\boldsymbol{a}_{\text{SRP}}$ denotes the acceleration induced by the SRP. $\boldsymbol{a}^{P_j}_{\text{3rd}}$ represents the third-body perturbation acceleration due to the $j$-th external body, which is expressed by:
\begin{equation}
    \boldsymbol{a}^{P_j}_{\text{3rd}} = -\mu_{P_j}\, \left( \dfrac{\boldsymbol{r}_{\text{EJ2000}} - \boldsymbol{r}^{P_j}_{\text{EJ2000}}}{\left\|\boldsymbol{r}^{P_j}_{\text{EJ2000}} - \boldsymbol{r}^{P_j}_{\text{EJ2000}}\right\|^3 } + \dfrac{\boldsymbol{r}^{P_j}_{\text{EJ2000}}}{\left\|\boldsymbol{r}^{P_j}_{\text{EJ2000}}\right\|^3} \right)
\label{eq:3rdBodyAcc}
\end{equation}
where $\boldsymbol{r}^{P_j}_{\text{EJ2000}}$ denotes the position vector of the $j$-th external body in the ECLIPJ2000 frame. In this study, the external bodies considered include the Sun, Mars, and Jupiter. The ephemerides of these celestial bodies, together with those of Didymos and Dimorphos, are retrieved from the latest NASA's DART mission SPICE kernels \citep{Chabot2024achievement}.\footnote{NASA's DART mission SPICE kernels are publicly available at: \url{https://naif.jpl.nasa.gov/pub/naif/pds/pds4/dart/dart_spice/spice_kernels/}} In addition, the post-impact physical parameters of Didymos system used in this study are listed in Table \ref{tab:DidymosSystemProperties}. The formulations of acceleration terms $\boldsymbol{a}^{D_i}_{\text{EJ2000}}$ and $\boldsymbol{a}_{\text{SRP}}$ are detailed in the following subsections. 

\begin{table}[htbp]
    \centering
    \caption{Post-impact physical parameters of Didymos system.}
    \label{tab:DidymosSystemProperties}
    \small
    \begin{tabular}{llll}
        \toprule\toprule
        Physical Parameters & Units & Values $\pm$ 1$\sigma$ \\
        \midrule

        System bulk density
        & kg/m$^3$
        & $2790 \pm 140$ \\

        Mutual centre separation 
        & km
        & $1.189 \pm 0.017$ \\

        Mutual orbital period 
        & hr
        & $11.3674 \pm 0.0004$ \\

        Mutual orbital eccentricity 
        & --
        & $0.0274 \pm 0.0015$ \\

        Didymos major semi-axis 
        & m
        & $394 \pm 11$ \\

        Dimorphos major semi-axis 
        & m
        & $96 \pm 6$ \\

        Geometric albedo
        & --
        & $0.15 \pm 0.02$ \\
        \bottomrule
    \end{tabular}

    \vspace{0.5em}
    \begin{minipage}{\linewidth}
        \footnotesize
        \textbf{Note.} The values of physical parameters listed in Table \ref{tab:DidymosSystemProperties} are taken from \cite{Naidu2024Orbital}, except for the geometric albedo, which is taken from \cite{Daly2023Successful}.
    \end{minipage}
\end{table}

\subsection{Gravitational acceleration of Didymos and Dimorphos}
{\label{C22:SubSection22}}


Considering the influence of the binary asteroids' irregular gravitational fields on the motion of nearby ejecta, the gravitational acceleration due to Didymos and Dimorphos is formulated using a distance-dependent piecewise model. When an ejecta particle is sufficiently far from an asteroid, the standard point-mass gravity model is adopted. In the vicinity of the asteroid, where the irregular body shape has non-negligible effect on the local gravitational field, the point-mass approximation is replaced by a polyhedron gravity model. In the polyhedron model, the shape of an asteroid is represented by a polyhedron composed of substantial planar triangular facets, which enables an accurate evaluation of the gravitational field associated with the asteroid's irregular mass distribution. In this work, the algorithm proposed by \citet{Werner1996Exterior} is employed. It provides closed-form formulae for the exterior gravitation due to a constant-density polyhedron, expressed in the corresponding asteroid body-fixed (BF) frame. Following this formulation, the gravitational acceleration due to Didymos or Dimorphos in its BF frame is given by: 
\begin{equation}
    \boldsymbol{a}_{poly,\text{BF}} = \nabla U_{poly,\text{BF}} = -G \sigma \sum_{e=1}^{n_e} \boldsymbol{E}_e \cdot \boldsymbol{r}_e \cdot L_e + G \sigma \sum_{f=1}^{n_f} \boldsymbol{F}_f \cdot \boldsymbol{r}_f \cdot \omega_f
	\label{eq:PolyhedronAcc_BodyFixed}
\end{equation}
where $G$ is the gravitational constant and $\sigma$ is the asteroid's bulk density. The quantities $n_e$ and $n_f$ denote the numbers of edges and faces of the polyhedron, respectively. The vector $\boldsymbol{r}_e$ is defined from an external point to an arbitrary point on edge $e$, while vector $\boldsymbol{r}_f$ is defined from an external point to an arbitrary point on face $f$. The dyads $\boldsymbol{E}_e$ and $\boldsymbol{F}_f$ are constructed from the normal vectors associated with edge $e$ and face $f$, respectively. The quantities $L_e$ and $\omega_f$ are the corresponding edge and face integration factors. Detailed expressions for $\boldsymbol{E}_e$, $\boldsymbol{F}_f$, $L_e$, and $\omega_f$ are provided in the relevant reference by \citet{Fu2024Dynamics}. In addition, the Laplacian of the polyhedron gravitational potential $U_{poly}$ is derived as:
\begin{equation}
    \nabla^2 U_{poly,\text{BF}} =-G \sigma \sum_{f=1}^{n_f} \omega_f 
                       = \left\{\begin{array}{cc}
    	0 & \text { outside } \\
    	-4 \pi G \sigma & \text { inside }
    				\end{array}\right.
	\label{eq:PolyhedronLaplacian_BodyFixed}
\end{equation}
This property provides a robust criterion for determining whether a point lies inside or outside the polyhedron. It is therefore utilised in this study to detect impacts of ejecta particles on the surfaces of Didymos and Dimorphos during their orbital evolution. Additionally, the up-to-date shape models of Didymos and Dimorphos, \textit{didymos\_g\_09309mm\_spc\_0000n00000\_v003.bds} and \textit{dimorphos\_g\_01940mm\_spc\_0000n00000\_v004.bds}, are adopted in this research. Both shape models are derived from DRACO and LICIACube images and enable polyhedron representations of the two asteroids for high-fidelity gravity field evaluation \citep{Daly2024Updated}.

Based on the polyhedron acceleration in Eq. (\ref{eq:PolyhedronAcc_BodyFixed}), the piecewise function of Didymos or Dimorphos gravitational acceleration in the ECLIPJ2000 frame is formulated as:
\begin{equation}
    \boldsymbol{a}^{D_i}_{\text{EJ2000}} = \left\{ \begin{array}{cc}
    \boldsymbol{M}^{D_i}_{\text{BFtoEJ2000}} \cdot \boldsymbol{a}^{D_i}_{poly,\text{BF}} & 
    d^{D_i} \leq d^{D_i}_{thresh} \\[2.0ex]
    -\mu_{D_i}\,\dfrac{\boldsymbol{r}_{\text{EJ2000}} - \boldsymbol{r}^{D_i}_{\text{EJ2000}}}{\left\|\boldsymbol{r}_{\text{EJ2000}} - \boldsymbol{r}^{D_i}_{\text{EJ2000}}\right\|^3 } &
    d^{D_i} > d^{D_i}_{thresh} 
    \end{array}\right.
    \label{eq:piecewiseAcc_Di}
\end{equation}
where $\boldsymbol{r}^{D_i}_{\text{EJ2000}}$ denotes the position vector of asteroid $D_i$ in the ECLIPJ2000 frame. $d^{D_i}=\left\|\boldsymbol{r}_{\text{EJ2000}} - \boldsymbol{r}^{D_i}_{\text{EJ2000}}\right\|$ is the distance between the ejecta particle and asteroid $D_i$'s barycentre. The threshold distance $d^{D_i}_{thresh}$ determines the transition between the polyhedron and point-mass models. Specifically, $d^{D_i}_{thresh}$ is defined as the distance beyond which the relative difference in acceleration magnitude between the two models is less than 0.1$\%$. Based on this criterion, the threshold distances for Didymos and Dimorphos are set to 16.0 km and 3.5 km, respectively. 

Additionally, in Eq. (\ref{eq:piecewiseAcc_Di}), $\boldsymbol{M}^{D_i}_{\text{BFtoEJ2000}}$ denotes the transformation matrix to transform the polyhedron acceleration $\boldsymbol{a}^{D_i}_{poly,\text{BF}}$ originally evaluated in asteroid $D_i$'s BF frame into the ECLIPJ2000 frame. The matrix $\boldsymbol{M}^{D_i}_{\text{BFtoEJ2000}}$ and its inverse matrix $\boldsymbol{M}^{D_i}_{\text{EJ2000toBF}}$ are time-dependent and can be readily retrieved from the DART SPICE kernels. To compute the polyhedron acceleration in Eq. (\ref{eq:piecewiseAcc_Di}), the relative position vector of an ejecta with respect to asteroid $D_i$ is first transformed from the ECLIPJ2000 frame to asteroid $D_i$'s BF frame using  $\boldsymbol{M}^{D_i}_{\text{EJ2000toBF}}$. The transformed position vector is then substituted into Eq. (\ref{eq:PolyhedronAcc_BodyFixed}) to evaluate $\boldsymbol{a}^{D_i}_{poly,\text{BF}}$, which subsequently enables the computation in Eq. (\ref{eq:piecewiseAcc_Di}).

\subsection{SRP acceleration with combined occultations}
{\label{C23:SubSection23}}

The SRP acceleration acting on an ejecta particle is approximated using the cannonball model \citep{Scheeres2016Orbital}. Under this approximation, the particle is treated as a sphere with a constant reflectivity and a constant projected cross-sectional area. In addition, the solar occultations caused by Didymos and Dimorphos are taken into account in this research. Although the occultations have not been widely considered in previous studies on ejecta evolution, they can influence the motion of small low-speed ejecta in the near-binary environment. For instance, the SRP acceleration acting on a spherical ejecta particle with a radius of \(1~\mathrm{mm}\) can reach the same order of magnitude as the gravitational acceleration due to Didymos and Dimorphos \citep{Fu2025Orbital}. Therefore, variations in the SRP acceleration, including its temporary reduction or complete suppression during occultation, may significantly affect the subsequent orbital evolution of such particles. The cannonball-model SRP acceleration incorporating the solar occultation is formulated as:
\begin{equation}
    \boldsymbol{a}_{\text{SRP}} =
    -\nu \frac{S_0}{c} C_r \frac{A}{m} \left(\frac{d_0}
    {\left\|
    \boldsymbol{r}^{S}_{\mathrm{EJ2000}} - \boldsymbol{r}_{\mathrm{EJ2000}}
    \right\|
    }\right)^2
    \frac{\boldsymbol{r}^{S}_{\mathrm{EJ2000}} - \boldsymbol{r}_{\mathrm{EJ2000}}}
    {
    \left\|
    \boldsymbol{r}^{S}_{\mathrm{EJ2000}} - \boldsymbol{r}_{\mathrm{EJ2000}}
    \right\|}
    \label{eq:SRP_cannonball_Occultation}
\end{equation}
where $S_0=$1367$~\mathrm{W/m^{2}}$ is the solar flux at \(1~\mathrm{AU}\), \(c\) is the speed of light, \(C_r=1+\rho_s\) is the reflectivity coefficient and $\rho_s$ is the geometric albedo, \(A\) is the projected cross-sectional area of the particle, \(m\) is its mass, and \(d_0=1~\mathrm{AU}\). Note that the density of a DART impact ejecta particle is set to 3500~kg$/$m$^3$ to match the grain density of L and LL chondrites \citep{Flynn2015Hypervelocity}, which are the best meteoric analogues for Didymos \citep{Dunn2013Mineralogies}. The SRP attenuation due to the solar occultation is modelled by the shadow function $\nu\in[0,1]$, whose formulation is detailed below.



Given the finite apparent size of the Sun, the shadow cast by asteroid $D_i$ is represented using a conical shadow model. The apparent angular radius of the Sun or asteroid $D_i$ is given by
\begin{equation}
    \alpha_{body} = \sin^{-1}\left(
    \frac{R_{body}}{\left\|
    \boldsymbol{r}^{body}_{\mathrm{EJ2000}} - \boldsymbol{r}_{\mathrm{EJ2000}}
    \right\|}
    \right),\;\; (body = S, D_i)
    \label{eq:angularRadius}
\end{equation}
where \(R_{body}\) denotes the effective radius of the body of interest. For Didymos and Dimorphos, \(R_{body}\) is selected as the major semi-axis of the corresponding asteroid, providing a conservative estimate of its occulting cross-section. The apparent angular separation between the centre of the solar disk and that of the occulting body, as viewed from an ejecta particle, is given by
\begin{equation}
    \theta_{D_i}
    =
    \cos^{-1}
    \left(\frac{
    \left(\boldsymbol{r}^{S}_{\mathrm{EJ2000}} - \boldsymbol{r}_{\mathrm{EJ2000}} \right)
    \cdot
    \left(\boldsymbol{r}^{D_i}_{\mathrm{EJ2000}} - \boldsymbol{r}_{\mathrm{EJ2000}} \right)
    }
    {\left\|\boldsymbol{r}^{S}_{\mathrm{EJ2000}} - \boldsymbol{r}_{\mathrm{EJ2000}} \right\|
    \left\|\boldsymbol{r}^{D_i}_{\mathrm{EJ2000}} - \boldsymbol{r}_{\mathrm{EJ2000}} \right\|}
    \right)
    \label{eq:angular_separation}
\end{equation}

Based on the geometry defined by $\alpha_{body}$ and $\theta_{D_i}$, the single-body shadow function \(\nu_{D_i}\) is formulated from the overlap between the solar disk and the disk of the occulting asteroid on the observer sky plane \citep{Montenbruck2000SatelliteOrbits}. If the two apparent disks do not overlap, the particle is fully illuminated and \(\nu_{D_i}=1\); if the occulting body completely covers the solar disk, the particle is in umbra and \(\nu_{D_i}=0\); otherwise, the particle is in penumbra and \(\nu_{D_i}\) is determined from the visible fraction of the solar disk. Specifically,
\begin{equation}
    \nu_{D_i}
    =
    \left\{
    \begin{array}{cl}
    1
    &
    \theta_{D_i} \geq \alpha_S+\alpha_{D_i}
    \\[1.2ex]
    0
    &
    \theta_{D_i} \leq \alpha_{D_i}-\alpha_S
    \\[1.2ex]
    1-\dfrac{A_{\mathrm{ov}}^{D_i}}{\pi \alpha_S^2}
    &
    \alpha_{D_i}-\alpha_S < \theta_{D_i} < \alpha_S+\alpha_{D_i} 
    \end{array}
    \right.
    \label{eq:singleBodyShadowFunction}
\end{equation}
where in the penumbra case, the angular overlap area $A_{\mathrm{ov}}^{D_i}$ is:
%
%
\begin{equation}
\begin{aligned}
    A_{\mathrm{ov}}^{D_i}
    ={}&
    \alpha_S^2
    \cos^{-1}
    \left(
    \frac{\theta_{D_i}^2+\alpha_S^2-\alpha_{D_i}^2}
         {2\theta_{D_i}\alpha_S}
    \right)
    +
    \alpha_{D_i}^2
    \cos^{-1}
    \left(
    \frac{\theta_{D_i}^2+\alpha_{D_i}^2-\alpha_S^2}
         {2\theta_{D_i}\alpha_{D_i}}
    \right)
    \\
    &-
    \frac{1}{2}
    \sqrt{
    4\theta_{D_i}^2\alpha_S^2
    -
    \left(\theta_{D_i}^2+\alpha_S^2-\alpha_{D_i}^2\right)^2
    } 
\end{aligned}
\label{eq:diskOverlapArea}
\end{equation}
The formulation in Eq. (\ref{eq:singleBodyShadowFunction}) provides a continuous transition between full sunlight, penumbra, and umbra for a single occulting body. Regarding the binary system, an ejecta particle can be occulted by Didymos, by Dimorphos, or by both asteroids. In this case, the shadow functions associated with Didymos and Dimorphos are first evaluated independently using Eq.~(\ref{eq:singleBodyShadowFunction}). Subsequently, the combined binary-system shadow function is approximated as
\begin{equation}
    \nu =
    \min\left(\nu_{D_1},\nu_{D_2}\right)
    \label{eq:combinedOccultationShadowFunction}
\end{equation}
so that the stronger of the two occultations determines the effective SRP attenuation. This compact treatment captures the dominant occultation effect in the binary environment, while avoiding the need to compute the full three-disk overlap between the Sun, Didymos, and Dimorphos on the observer sky plane. It is noteworthy that Eq.~(\ref{eq:combinedOccultationShadowFunction}) does not explicitly account for the case in which Didymos and Dimorphos simultaneously cover different non-overlapping portions of the solar disk. Nevertheless, it provides a computationally efficient approximation suitable for large-scale ejecta propagation.

\section{Initial condition modelling for DART ejecta}
{\label{C3:Section3}}
This section presents the methodology for generating the initial conditions of ejecta particles produced by the DART impact. The initial ejection speeds are assigned according to a velocity-size distribution, which relates the spherical particle's initial speed to its radius. The particle directions are then prescribed using simulated ejecta cone geometries, thereby defining the complete initial velocity vectors. Note that both initial speed assignment and cone-based direction sampling are performed in the Dimorphos BF frame. The resulting initial conditions must be transformed into the ECLIPJ2000 frame before propagation. 

\subsection{Initial speed assignment}
{\label{C31:SubSection31}}

According to the impact scaling theory \citep{Holsapple2012Momentum}, the initial ejection speed is expected to be particle-size dependent, with smaller particles generally characterised by higher velocities than larger fragments. In this study, the velocity--size distribution proposed by \citet{Ferrari2025Morphology} is adopted to sample the initial ejection speeds of DART impact ejecta of different sizes. This distribution characterises the statistical relationship between particle radius and initial ejection speed, providing a physically motivated means to introduce size-dependent dispersion into the ejecta initial conditions. In \citet{Ferrari2025Morphology}, this formulation was shown to be critical for reproducing key morphological DART ejecta features observed in both HST images and associated synthetic images. This agreement supports its applicability to generate physically representative initial conditions for the ejecta population considered in this study.

Following the adopted velocity--size distribution, the initial ejection speed $v_0$ of a particle at the impact epoch $t_0$ is modelled as the sum of the escape velocity from the impact location on Dimorphos, $v_{esc}$, and a size-dependent ejection speed term:
\begin{equation}
    v_0 = v_{esc} + \rho v_{radius}^{max}
    \label{eq:VelocitySizeSpeed}
\end{equation}
where $\rho$ is a stochastic coefficient sampled from the standard uniform distribution $\mathcal{U}(0,1)$. The quantity $v_{radius}^{max}$ denotes the maximum size-dependent ejection speed assigned to a particle of radius $r$ regardless of $v_{esc}$, and is given by:
\begin{equation}
    v_{radius}^{max} = \kappa r_\text{ej}^{-\gamma}
    \label{eq:VelocitySizeMax}
\end{equation}
where $\kappa$ and $\gamma$ are the coefficient and exponent of the velocity--size distribution, respectively, and $r_\text{ej}$ is expressed in metres. The escape velocity of Dimorphos $v_{esc}$ is approximately 0.09$~\mathrm{m/s}$. The exponent $\gamma$ is set to 0.4, which is consistent with the value reported by \citet{Ferrari2025Morphology}. The coefficient $\kappa$ is selected so that $v_{radius}^{max}$ of a spherical particle with a radius of $10^{-6}$~m can reach 500$~\mathrm{m/s}$, corresponding to the approximate maximum ejecta speed inferred from LICIACube images \citep{Dotto2024Dimorphos}. Under the adopted velocity--size distribution, the maximum initial ejection speed of a 1~cm-radius particle is approximately 17.5$~\mathrm{m/s}$. 
\begin{figure}[ht!]
\centering
\includegraphics[width=0.84\hsize]{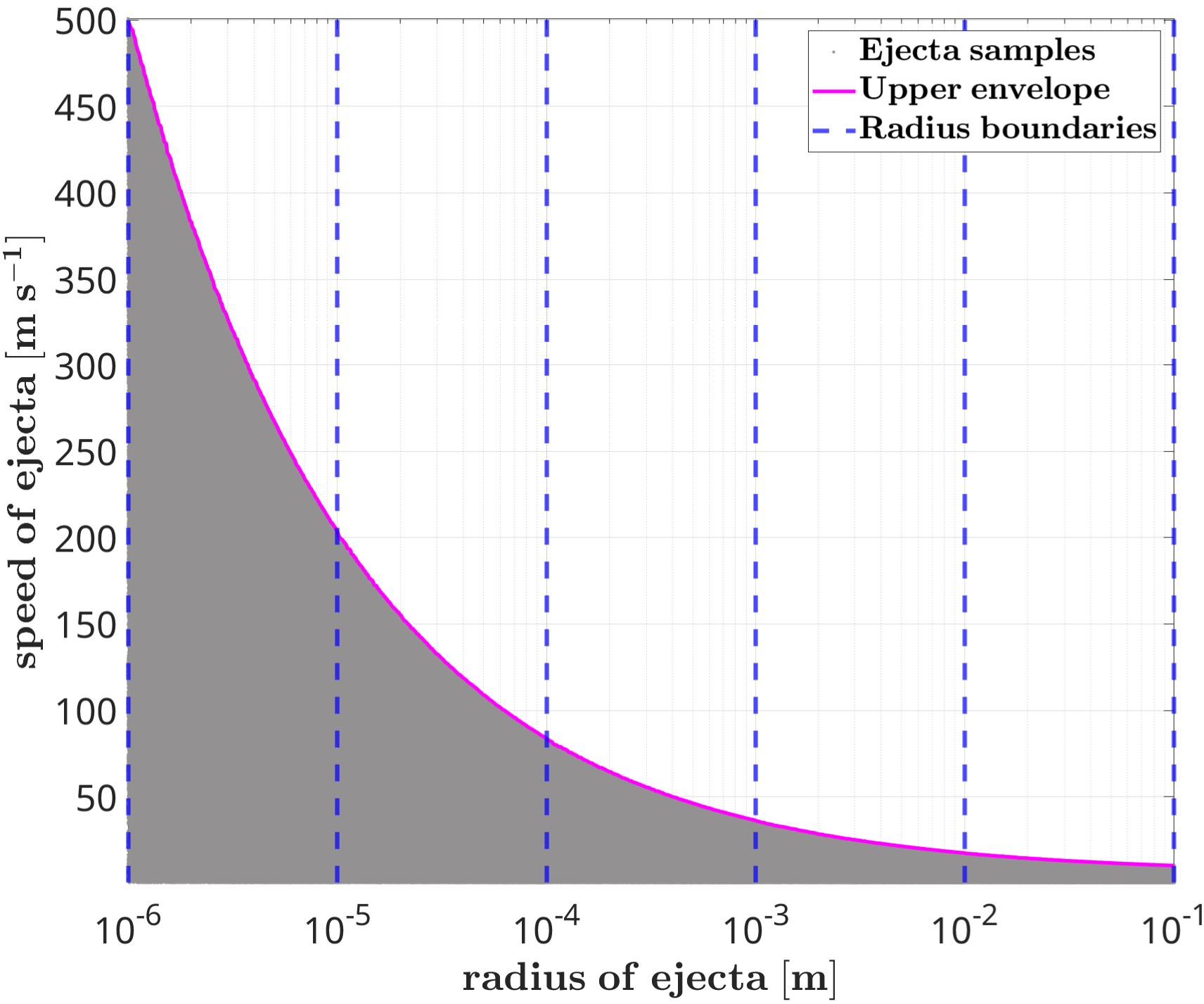}
    \caption{Speed--radius distribution of DART impact ejecta samples. The samples are classified into five size groups whose boundaries are marked by dashed blue vertical lines. Its upper envelope represents the maximum size-dependent ejection speed calculated from the adopted velocity--size distribution.}
    \label{fig:SpeedRadiusDistribution}
\end{figure}
\begin{figure*}[ht!]
    \centering
    \includegraphics[width=0.246\hsize]{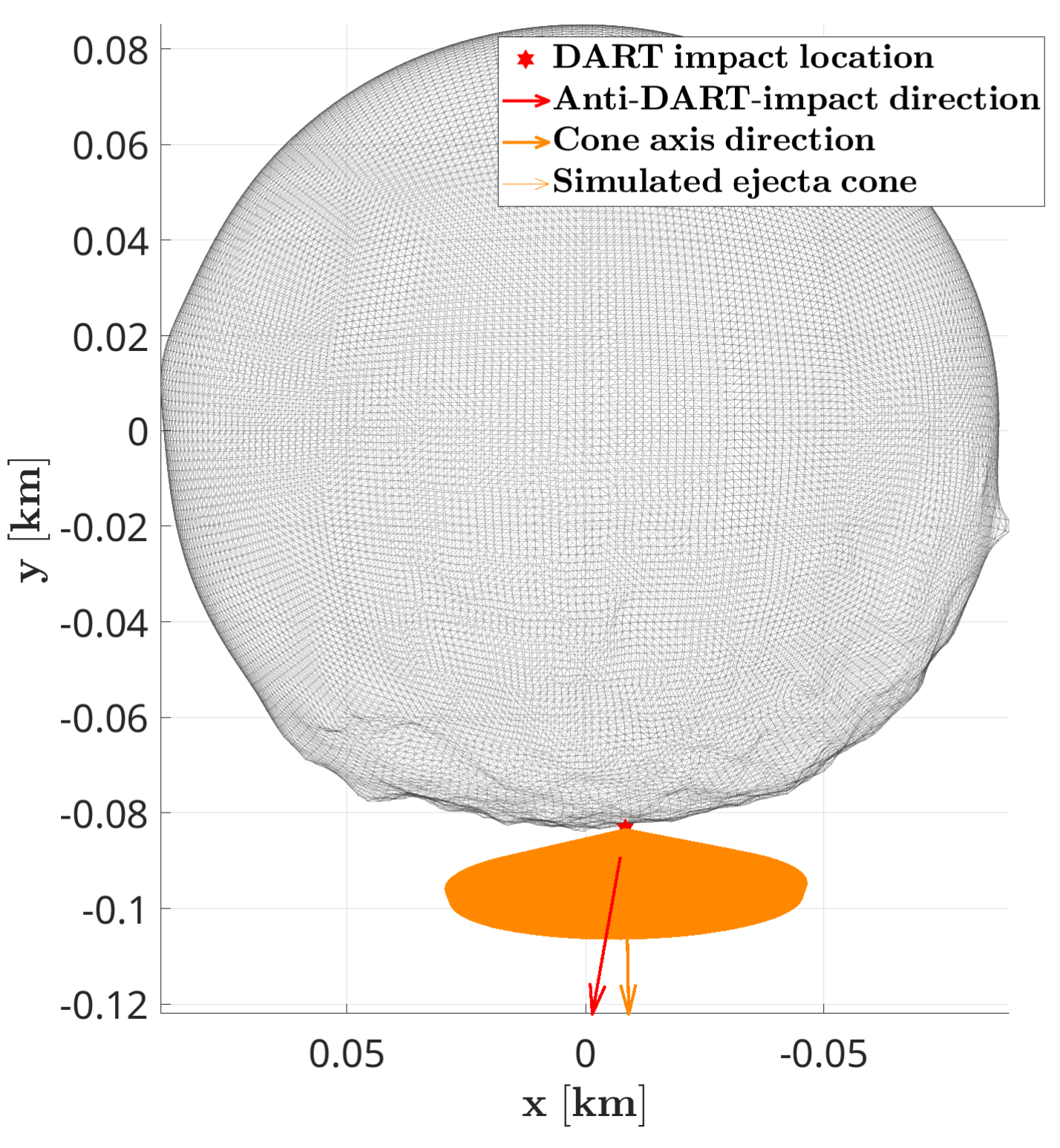}
    \includegraphics[width=0.246\hsize]{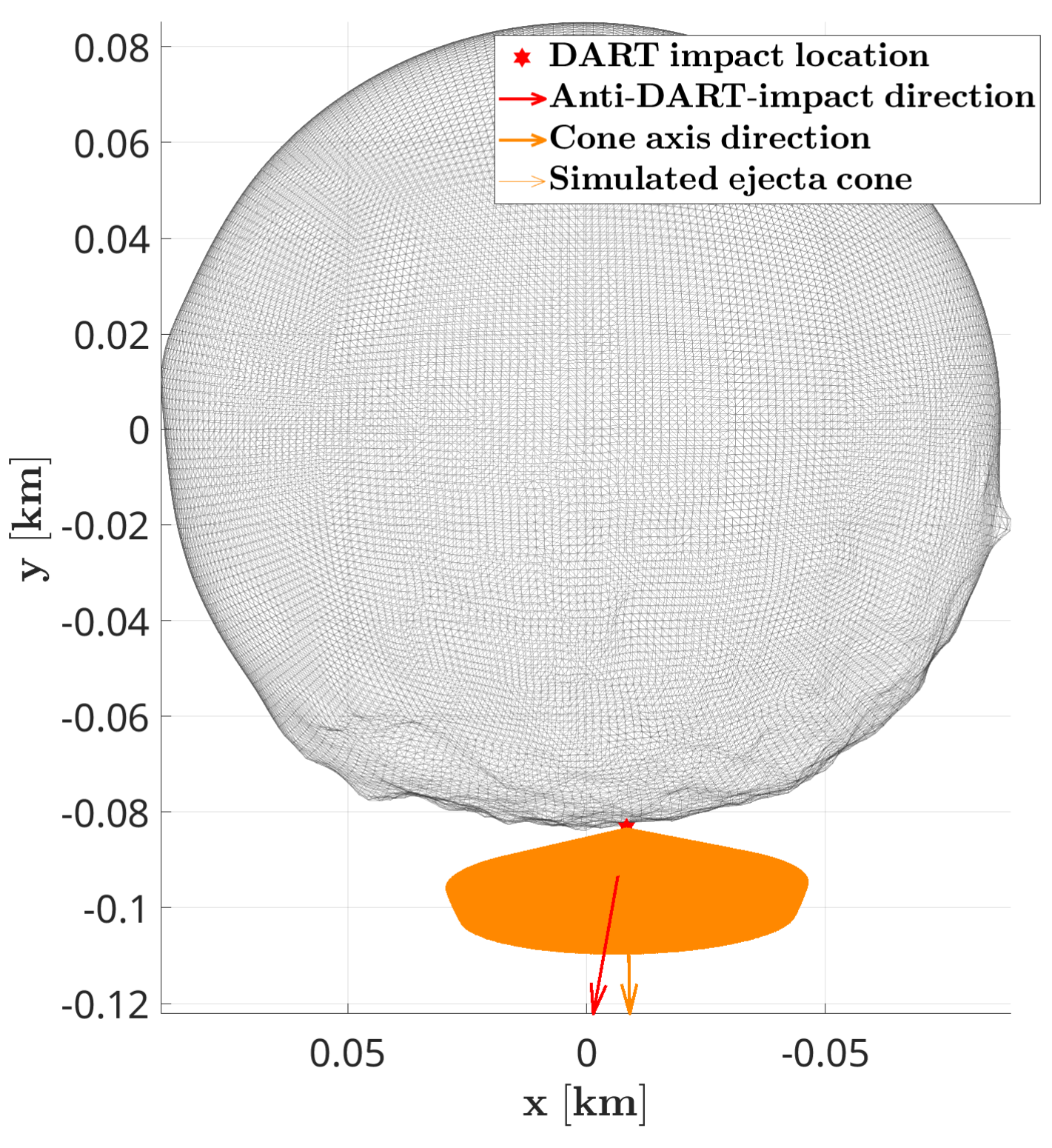}
    \includegraphics[width=0.246\hsize]{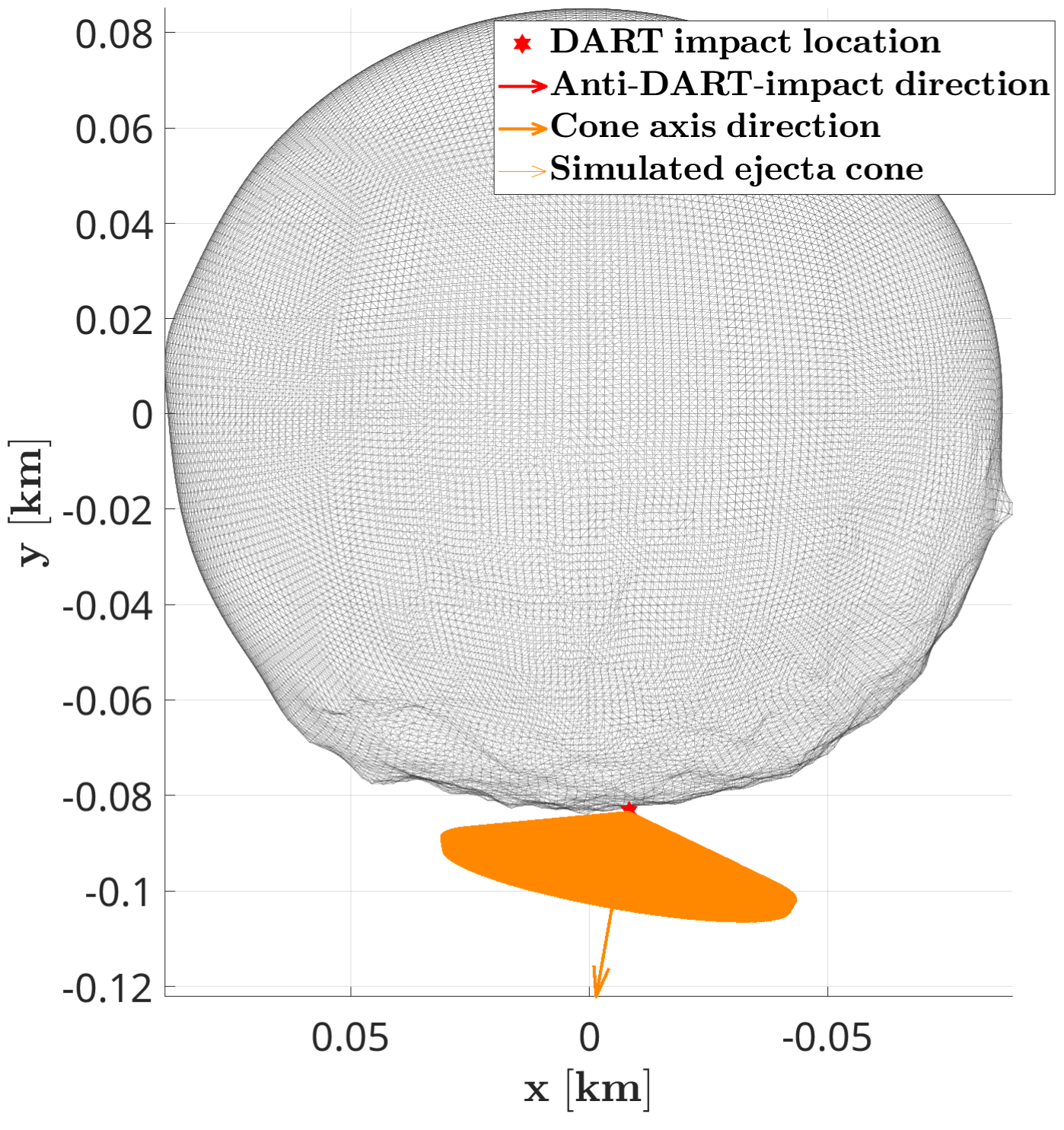}
    \includegraphics[width=0.246\hsize]{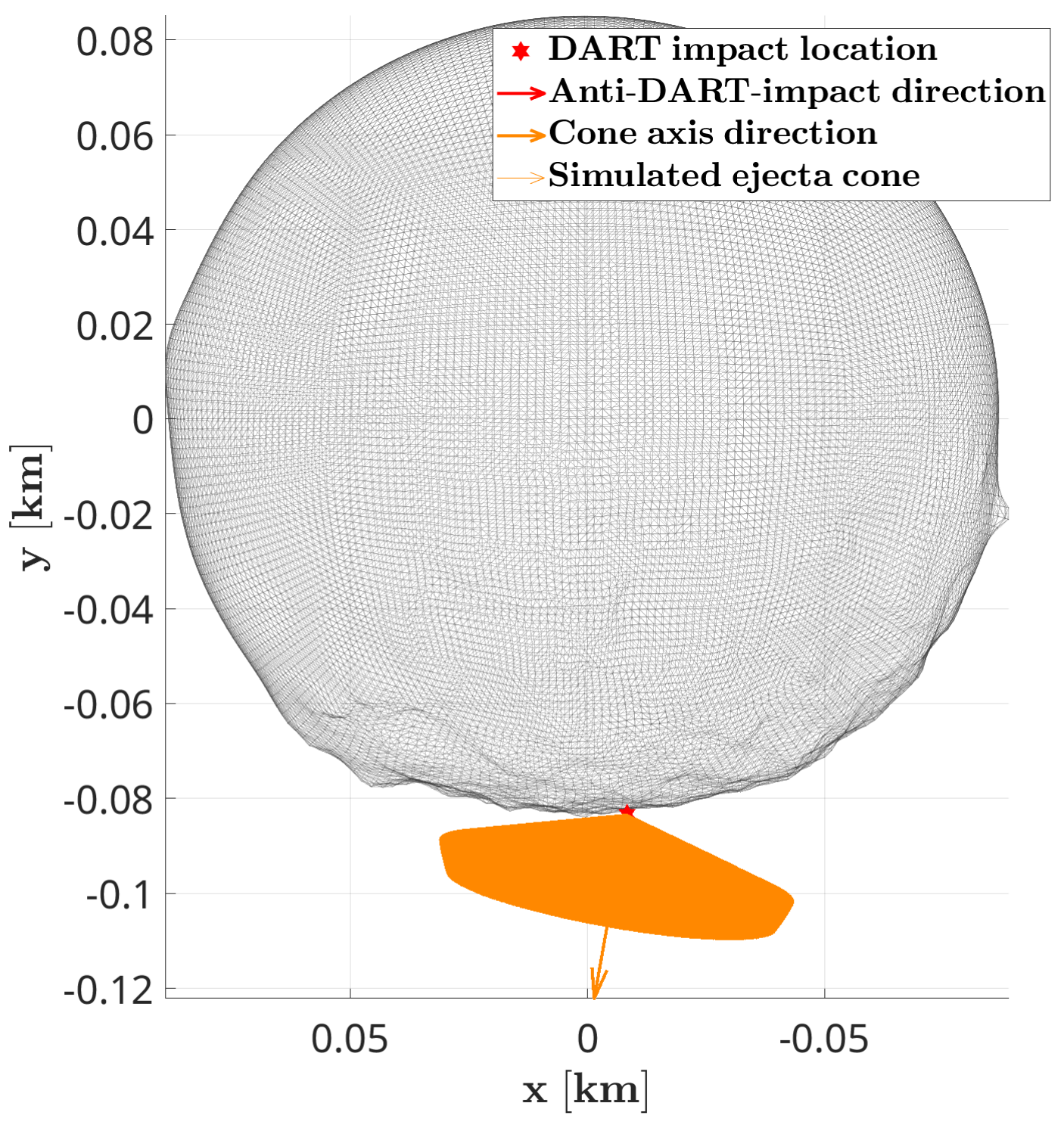}
    \caption{Four ejecta cone geometries. From left to right, the four panels correspond to cone geometries defined by 1) (RA, DEC) = $(140^\circ, 20^\circ)$ with $(\theta_{\min}, \theta_{\max})=(68^\circ, 72^\circ)$; and 2) (RA, DEC) = $(140^\circ, 20^\circ)$ with $(\theta_{\min}, \theta_{\max})=(62^\circ, 72^\circ)$; 3) anti-DART-impact direction with $(\theta_{\min}, \theta_{\max})=(68^\circ, 72^\circ)$; 4) anti-DART-impact direction with $(\theta_{\min}, \theta_{\max})=(62^\circ, 72^\circ)$;. Each simulated ejecta cone is constructed from  $10^6$ sample unit ejection vectors. Note that the cones are viewed from the south-pole direction of Dimorphos, which is consistent with similar illustrations in previous works.}
    \label{fig:EjectaConeGeometries}
\end{figure*}

Fig.~\ref{fig:SpeedRadiusDistribution} shows the speed--radius distribution of the ejecta samples used in this study. The ejecta particles are divided into five size groups defined by $r_\text{ej}=[10^{-(i+1)}, 10^{-i})$~m, with $i=1,...,5$. The boundaries of these groups are indicated by the dashed blue vertical lines in this figure. Each size group contains $10^6$ samples uniformly distributed within the associated radius range, and each sample is represented by a grey point. The upper envelope of the grey area corresponds to the maximum size-dependent ejection speed determined by the adopted velocity--size distribution.

\subsection{Cone-based direction sampling}
{\label{C32:SubSection32}}

The initial ejection direction of each particle is sampled within a prescribed ejecta cone defined in the Dimorphos BF frame. The ejecta cone geometry is specified by two factors, the cone-axis direction and the range of admissible half-cone angles measured from the cone axis. Let $\boldsymbol{e}_{\mathrm{axis}}$ denote the unit vector along the cone axis, and $\theta$ denote the half-cone angle. The ejecta directions are sampled over a conical shell bounded by two half-cone angles, $\theta_{\min}$ and $ \theta_{\max}$. 

To generate an isotropic distribution of directions over the conical shell, the azimuth angle $\phi$ is sampled uniformly from \([0,2\pi)\), while \(\cos\theta\) is sampled uniformly over the interval defined by the conical shell boundaries, i.e., 
\begin{equation}
    \phi \sim \mathcal{U}(0,2\pi),\qquad \cos\theta \sim \mathcal{U}\left(\cos\theta_{\max}, \cos\theta_{\min}\right).
    \label{eq:HalfConeAngleDistribution}
\end{equation}
This sampling strategy ensures a uniform distribution over the solid angle within the prescribed range of half-cone angles. An orthonormal basis $\{\boldsymbol{e}_1,\boldsymbol{e}_2,\boldsymbol{e}_{\mathrm{axis}}\}$ is then constructed around the cone axis. Two transverse unit vectors $\boldsymbol{e}_1$ and $\boldsymbol{e}_2$ span the plane normal to $\boldsymbol{e}_{\mathrm{axis}}$, satisfying:
\begin{equation}
    \boldsymbol{e}_1 \cdot \boldsymbol{e}_{\mathrm{axis}} = 0,
    \qquad
    \boldsymbol{e}_2 \cdot \boldsymbol{e}_{\mathrm{axis}} = 0,
    \qquad
    \boldsymbol{e}_1 \cdot \boldsymbol{e}_2 = 0
    \label{eq:BasisRelation}
\end{equation}
The unit ejection direction vector, $\boldsymbol{u}_{\mathrm{ej}}$, is then expressed as:
\begin{equation}
    \boldsymbol{u}_{\mathrm{ej}}
    =
    \sin\theta\cos\phi\,\boldsymbol{e}_1
    +
    \sin\theta\sin\phi\,\boldsymbol{e}_2
    +
    \cos\theta\,\boldsymbol{e}_{\mathrm{axis}} .
    \label{eq:UnitEjectionDirection}
\end{equation}
By construction, \(\|\boldsymbol{u}_{\mathrm{ej}}\|=1\), and the angle between \(\boldsymbol{u}_{\mathrm{ej}}\) and the cone-axis vector $\boldsymbol{e}_{\mathrm{axis}}$ satisfies $\theta_{\min} \leq \cos^{-1} ( \boldsymbol{u}_{\mathrm{ej}}\cdot\boldsymbol{e}_{\mathrm{axis}}) \leq \theta_{\max}$. Subsequently, the initial ejection speed $v_0$ generated using the velocity--size distribution described in Section \ref{C31:SubSection31} is assigned to a unit ejection direction vector, giving a particle's initial velocity vector $\boldsymbol{v}_0$:
\begin{equation}
    \boldsymbol{v}_0 = v_0 \boldsymbol{u}_{\mathrm{ej}}
    \label{eq:IntialVelocityVector}
\end{equation}
The initial velocity vector $\boldsymbol{v}_0$, together with the initial position vector defined by the DART impact location, completes the initial state of an ejecta particle. The DART impact location has been accurately determined as 8.84$^\circ$S and 264.30$^\circ$E in the Dimorphos BF frame \citep{Daly2023Successful}. The corresponding position vector can be retrieved from NASA's DART SPICE kernels. 

In contrast, the geometry of the DART ejecta cone remains subject to uncertainty. Regarding the cone-axis direction, multiple studies reported that the cone axis is aligned with the anti-DART-impact direction, e.g., in \cite{Li2023Ejecta} and \cite{Ferrari2025Morphology}. However, recent analyses based on LICIACube images suggest that the ejecta cone axis is not strictly aligned with the incoming direction of the DART spacecraft, but is instead better represented by a direction specified in right ascension (RA) and declination (DEC) in the J2000 frame. For instance, \cite{Dotto2024Dimorphos} reported an updated cone axis direction of (RA, DEC) = ($137^{+8}_{-9}$$^{\circ}$, $19^{+10}_{-12}$$^{\circ}$). Another close estimate of ($141\pm4$$^{\circ}$, $20\pm8$$^{\circ}$) was reported by \cite{Hirabayashi2025elliptical}. Uncertainty also exists in the opening angle of the ejecta cone. Based on LICIACube images,  \cite{Dotto2024Dimorphos} reported an aperture angle of $140\pm4^{\circ}$, corresponding to a half-cone angle range of $(\theta_{\min}, \theta_{\max})=(68^\circ, 72^\circ)$. By further combining constraints from HST images, \cite{Hirabayashi2025elliptical} reported a cone wide angle of $133\pm9^{\circ}$, corresponding to $(\theta_{\min}, \theta_{\max})=(62^\circ, 71^\circ)$.

To account for the uncertainties summarised above, four ejecta-cone geometries are considered in this study. They are constructed from two possible cone axis directions, (RA, DEC) = $(140^\circ, 20^\circ)$ and the anti-DART-impact direction, and two possible half-cone angle ranges, $(\theta_{\min}, \theta_{\max})=(68^\circ, 72^\circ)$ and $(62^\circ, 72^\circ)$. The four resulting cone geometries are illustrated in Fig.~\ref{fig:EjectaConeGeometries}. Each simulated ejecta cone in this figure is constructed from $10^6$ unit ejection direction vectors. The application of the four cone geometries enables the influence of both the cone-axis direction and conical-shell width on the subsequent evolution of ejecta particles to be investigated.

\section{Evolution simulation of DART ejecta}
{\label{C4:Section4}}
This section presents the statstics of the DART ejecta evolution simulations. The numerical setup and simulation configuration, based on the dynamical model in Sect.~\ref{C2:Section2}, are first introduced. Subsequently, the initial conditions generated by the methodology described in Sect.~\ref{C3:Section3} are propagated, and the resulting dynamical evolution of the ejecta particles is analysed in detail.

\subsection{Numerical setup and simulation configurations}
{\label{C41:SubSection41}}

To improve the numerical robustness and conditioning of the integration, the ejecta dynamics are normalised using characteristic scales associated with the Didymos binary system. The characteristic gravitational parameter is taken as the total gravitational parameter of the system, $\mu^* = \mu_{D_1} + \mu_{D_2}$, while the characteristic length is selected as the binary's semi-major axis, $ L^* = a_{\mathrm{binary}}$. The corresponding characteristic time and velocity are then defined as $T^* = \sqrt{{L^*}^3/\mu^*}$ and $ V^* = {L^*}/{T^*}$, respectively. These normalisation units bring the propagated position and velocity variables to comparable non-dimensional magnitudes, thereby improving the numerical conditioning of the integration.

After normalisation, the ejecta dynamics are integrated using an adaptive-step Runge--Kutta--Fehlberg 7(8) embedded scheme (RKF78) implemented in the C++ Boost library \citep{Schaling2011boost}. The absolute and relative tolerances of the integrator are both set to 1.0$\times$10$^{-14}$ to minimise numerical error accumulation and maintain robust propagation during close-encounters. The initial and minimum integration step sizes are also set to 1.0$\times$10$^{-14}$ in non-dimensional time units, allowing the integrator to resolve rapid dynamical variations near Didymos or Dimorphos. The integrator framework also includes a prescribed event detection and event-state storage, enabling efficient identification and recording of key outcomes during the ejecta evolution simulations.

All modelling components, including gravitational accelerations, third-body perturbations, SRP with combined occultations, event detection and storage, and normalised integration scheme, are incorporated into the $N$-body propagator \textit{goNEAR}. This propagator has previously been successfully applied to the navigation analyses for JAXA's Hayabusa2 mission \citep{Soldini2020Hayabusa2,Soldini2022Superior}, which provides confidence in its eligibility for the large-scale ejecta evolution simulations conducted in this study.

The five ejecta size groups defined in Sect. \ref{C31:SubSection31}, denoted by $r_{\text{ej}}=[10^{-(i+1)}, 10^{-i})$~m with $(i$ = $1,...,5)$, are each assigned to the four candidate ejecta cone geometries described in Sect. \ref{C32:SubSection32}. Each ejecta size group contains 1 million samples. Therefore, combining the five size groups with the four ejecta cone geometries yields a total of 20 million initial conditions, which subsequently corresponds to 20 million individual ejecta evolution simulations. All simulations are initialised at the DART impact epoch and then propagated in the binary-barycentric ECLIPJ2000 frame using the dynamical model in Sect.~\ref{C2:Section2}. Each trajectory is propagated for approximately two years, around 731 days, unless a prescribed terminal event, such as impacting with Didymos or Dimorphos, occurs earlier.

\subsection{Statistical analysis of ejecta particle fates}
{\label{C42:SubSection42}}

The fates of ejecta particles in the 20 million independent simulations are classified into four categories. Apart from impacting on the surface of polyhedral Didymos or Dimorphos, which can be detected using in the inside--outside criterion given by Eq.~(\ref{eq:PolyhedronLaplacian_BodyFixed}), a third outcome is to escape from the gravitational dominance of the binary system. Using the reference pre-impact mass of 5.3$\times$10$^{11}$kg reported by \cite{Naidu2024Orbital}, the system's Hill's radius is estimated to vary approximately within the range 70--155 km over its heliocentric orbital period. In this study, a more conservative escape threshold of 300 km from the binary barycentre is adopted to ensure that particles classified as escaped are no longer primarily governed by the binary's gravitation. Based on these impact and escape criteria, the fourth outcome is defined as particles that neither impact Didymos or Dimorphos nor escape within the prescribed propagation time span. The resulting fate statistics for all simulated particles are summarised in Table~\ref{tab:EjectaFateStatistics}. 

\begin{table*}[hbtp]
    \centering
    \caption{Fate statistics of ejecta particles for different size groups and ejecta-cone geometries.}
    \label{tab:EjectaFateStatistics}
    \small
    \renewcommand{\arraystretch}{1.2}
    \begin{tabular}{c c c cccc}
        \toprule \toprule
        \multicolumn{1}{c}{\textbf{Size group}} &
        \multicolumn{2}{c}{\textbf{Ejecta-cone geometry}} &
        \multicolumn{4}{c}{\textbf{Fate of ejecta particles}} \\
        \cmidrule(lr){1-1}
        \cmidrule(lr){2-3}
        \cmidrule(lr){4-7}
        \makecell{Ejecta radius \\ $r_{\mathrm{ej}}$ [m]} &
        \makecell{Cone-axis\\direction} &
        \makecell{Half-cone angle\\ $\theta$ [deg]} &
        \makecell{No escape\\\& no impact} &
        \makecell{Impact on\\Didymos} &
        \makecell{Impact on\\Dimorphos} &
        \makecell{Escape from\\D\&D system} \\
        \midrule

        \multirow{4}{*}{[$10^{-2}$, $10^{-1}$)}
        & \cellcolor{lightred}{updated}
        & \cellcolor{lightred}{[68, 72]}
        & \cellcolor{lightred}{0.012\%}
        & \cellcolor{lightred}{0.348\%}
        & \cellcolor{lightred}{0.222\%}
        & \cellcolor{lightred}{99.418\%} \\

        & \cellcolor{lightyellow}{updated}
        & \cellcolor{lightyellow}{[62, 72]}
        & \cellcolor{lightyellow}{0.009\%}
        & \cellcolor{lightyellow}{0.323\%}
        & \cellcolor{lightyellow}{0.207\%}
        & \cellcolor{lightyellow}{99.461\%} \\

        & \cellcolor{lightgreen}{anti-DART}
        & \cellcolor{lightgreen}{[68, 72]}
        & \cellcolor{lightgreen}{0.008\%}
        & \cellcolor{lightgreen}{5.330\%}
        & \cellcolor{lightgreen}{0.203\%}
        & \cellcolor{lightgreen}{94.459\%} \\

        & \cellcolor{lightblue}{anti-DART}
        & \cellcolor{lightblue}{[62, 72]}
        & \cellcolor{lightblue}{0.009\%}
        & \cellcolor{lightblue}{3.033\%}
        & \cellcolor{lightblue}{0.196\%}
        & \cellcolor{lightblue}{96.762\%} \\
        \midrule

        \multirow{4}{*}{[$10^{-3}$, $10^{-2}$)}
        & \cellcolor{lightred}{updated}
        & \cellcolor{lightred}{[68, 72]}
        & \cellcolor{lightred}{0.000\%}
        & \cellcolor{lightred}{0.229\%}
        & \cellcolor{lightred}{0.041\%}
        & \cellcolor{lightred}{99.730\%} \\

        & \cellcolor{lightyellow}{updated}
        & \cellcolor{lightyellow}{[62, 72]}
        & \cellcolor{lightyellow}{0.000\%}
        & \cellcolor{lightyellow}{0.207\%}
        & \cellcolor{lightyellow}{0.039\%}
        & \cellcolor{lightyellow}{99.754\%} \\

        & \cellcolor{lightgreen}{anti-DART}
        & \cellcolor{lightgreen}{[68, 72]}
        & \cellcolor{lightgreen}{0.000\%}
        & \cellcolor{lightgreen}{5.816\%}
        & \cellcolor{lightgreen}{0.036\%}
        & \cellcolor{lightgreen}{94.148\%} \\

        & \cellcolor{lightblue}{anti-DART}
        & \cellcolor{lightblue}{[62, 72]}
        & \cellcolor{lightblue}{0.000\%}
        & \cellcolor{lightblue}{3.377\%}
        & \cellcolor{lightblue}{0.036\%}
        & \cellcolor{lightblue}{96.587\%} \\
        \midrule

        \multirow{4}{*}{[$10^{-4}$, $10^{-3}$)}
        & \cellcolor{lightred}{updated}
        & \cellcolor{lightred}{[68, 72]}
        & \cellcolor{lightred}{0.000\%}
        & \cellcolor{lightred}{0.042\%}
        & \cellcolor{lightred}{0.004\%}
        & \cellcolor{lightred}{99.954\%} \\

        & \cellcolor{lightyellow}{updated}
        & \cellcolor{lightyellow}{[62, 72]}
        & \cellcolor{lightyellow}{0.000\%}
        & \cellcolor{lightyellow}{0.039\%}
        & \cellcolor{lightyellow}{0.003\%}
        & \cellcolor{lightyellow}{99.958\%} \\

        & \cellcolor{lightgreen}{anti-DART}
        & \cellcolor{lightgreen}{[68, 72]}
        & \cellcolor{lightgreen}{0.000\%}
        & \cellcolor{lightgreen}{6.054\%}
        & \cellcolor{lightgreen}{0.004\%}
        & \cellcolor{lightgreen}{93.942\%} \\

        & \cellcolor{lightblue}{anti-DART}
        & \cellcolor{lightblue}{[62, 72]}
        & \cellcolor{lightblue}{0.000\%}
        & \cellcolor{lightblue}{3.639\%}
        & \cellcolor{lightblue}{0.003\%}
        & \cellcolor{lightblue}{96.358\%} \\
        \midrule

        \multirow{4}{*}{[$10^{-5}$, $10^{-4}$)}
        & \cellcolor{lightred}{updated}
        & \cellcolor{lightred}{[68, 72]}
        & \cellcolor{lightred}{0.000\%}
        & \cellcolor{lightred}{0.008\%}
        & \cellcolor{lightred}{0.001\%}
        & \cellcolor{lightred}{99.991\%} \\

        & \cellcolor{lightyellow}{updated}
        & \cellcolor{lightyellow}{[62, 72]}
        & \cellcolor{lightyellow}{0.000\%}
        & \cellcolor{lightyellow}{0.007\%}
        & \cellcolor{lightyellow}{0.002\%}
        & \cellcolor{lightyellow}{99.991\%} \\

        & \cellcolor{lightgreen}{anti-DART}
        & \cellcolor{lightgreen}{[68, 72]}
        & \cellcolor{lightgreen}{0.000\%}
        & \cellcolor{lightgreen}{6.283\%}
        & \cellcolor{lightgreen}{0.002\%}
        & \cellcolor{lightgreen}{93.715\%} \\

        & \cellcolor{lightblue}{anti-DART}
        & \cellcolor{lightblue}{[62, 72]}
        & \cellcolor{lightblue}{0.000\%}
        & \cellcolor{lightblue}{3.869\%}
        & \cellcolor{lightblue}{0.002\%}
        & \cellcolor{lightblue}{96.129\%} \\
        \midrule

        \multirow{4}{*}{[$10^{-6}$, $10^{-5}$)}
        & \cellcolor{lightred}{updated}
        & \cellcolor{lightred}{[68, 72]}
        & \cellcolor{lightred}{0.000\%}
        & \cellcolor{lightred}{0.035\%}
        & \cellcolor{lightred}{0.007\%}
        & \cellcolor{lightred}{99.957\%} \\

        & \cellcolor{lightyellow}{updated}
        & \cellcolor{lightyellow}{[62, 72]}
        & \cellcolor{lightyellow}{0.000\%}
        & \cellcolor{lightyellow}{0.039\%}
        & \cellcolor{lightyellow}{0.006\%}
        & \cellcolor{lightyellow}{99.956\%} \\

        & \cellcolor{lightgreen}{anti-DART}
        & \cellcolor{lightgreen}{[68, 72]}
        & \cellcolor{lightgreen}{0.000\%}
        & \cellcolor{lightgreen}{6.477\%}
        & \cellcolor{lightgreen}{0.007\%}
        & \cellcolor{lightgreen}{93.516\%} \\

        & \cellcolor{lightblue}{anti-DART}
        & \cellcolor{lightblue}{[62, 72]}
        & \cellcolor{lightblue}{0.000\%}
        & \cellcolor{lightblue}{4.018\%}
        & \cellcolor{lightblue}{0.007\%}
        & \cellcolor{lightblue}{95.975\%} \\

        \bottomrule
    \end{tabular}
\end{table*}

In Table~\ref{tab:EjectaFateStatistics}, the fate statistics are separated into 20 cases belonging to five ejecta size groups. In each group, the associated four candidate ejecta-cone geometries are highlighted using distinct background colours to facilitate comparison. For notational convenience, the anti-DART-impact direction is denoted as ``anti-DART'', whereas the cone-axis direction specified by (RA, DEC) = (140$^{\circ}$, 20$^{\circ}$) is denoted as ``updated''. The major findings from the fate statistics are summarised from four perspectives as follows.

\vspace{0.5em}
\noindent{\sffamily\itshape -- Overall trend} \par
\vspace{0.3em}

The fate statistics in Table~\ref{tab:EjectaFateStatistics} show that escape from the Didymos binary system is the dominant outcome for all ejecta size groups and cone geometries. In each case, more than 93.5\% of the simulated particles escape from the binary system within the prescribed propagation time span. This high escape fraction suggests that the majority of simulated particles are removed from the near-binary environment rather than remaining gravitationally bound to the system. Such behaviour is qualitatively consistent with the HST observations reported in \cite{Li2023Ejecta}, which revealed long ejecta tails that were still clearly visible approximately 15 days after impact. Although the observed brightness distribution depends on particle size, scattering properties, and viewing geometry, the large simulated escape fraction provides a physically plausible mechanism for sustaining extended ejecta structures over the observed timescale.

In addition, among the two impact outcomes, impacts on Didymos are consistently more frequent than impacts on Dimorphos. This contrast becomes particularly pronounced for cases associated with the ``anti-DART'' cone-axis direction. For instance, in the smallest size group, the Didymos impact fraction reaches \(6.477\%\) for the \([68^\circ,72^\circ]\) cone, whereas the corresponding Dimorphos impact fraction is only \(0.007\%\). Although the combined impact fraction remains below 7\% in all cases, this large contrast indicates that impact events are dominated by impact onto the primary asteroid, rather than by accretion onto Dimorphos. This behaviour should be associated with the specific geometry of the candidate ejection cones, the instantaneous binary-system configuration at the impact epoch, and the larger size and stronger gravitational influence of Didymos. 

\vspace{0.5em}
\noindent{\sffamily\itshape -- Dependence on particle size} \par
\vspace{0.3em}

The dependence on particle size is comparatively weak for the overall escape outcome, but some systematic trends can still be identified. For the ``updated'' cone-axis direction, the Didymos and Dimorphos impact fractions generally decrease from the largest particles to intermediate and smaller particles, while the escape fraction approaches nearly \(100\%\). This is consistent with the stronger influence of SRP on smaller particles, which promotes the removal of particles from the binary environment. However, for the smallest size group, the Didymos impact fraction slightly increases again under the ``updated'' direction, suggesting that the combined effects of SRP and binary gravity may introduce non-monotonic behaviour of the smallest ejecta.

However, it is noteworthy that the fate category in which particles neither escape nor impact either asteroid is identified only for the largest size group under the sampled cone geometries. Its fraction remains nearly insensitive to the cone-axis direction and half-cone angle range, staying at approximately 0.010\%. Furthermore, these ejecta fragments are characterised by an initial speed of no larger than 5 m/s. This suggests that long-term remaining in the near-binary environment is possible, but only for a very limited subset of relatively large low-mobility ejecta fragments. This behaviour is qualitatively consistent with the HST observations reported by \cite{Jewitt2023Dimorphos}, where a population of low-speed co-moving boulders was identified in the vicinity of the Didymos system approximately three months after impact. Although the simulated non-escaping fragments represent only a minute fraction of the ejecta population, their preferential occurrence in the largest size group supports the interpretation that larger fragments are more likely to remain dynamically associated with the binary system over extended timescales. Therefore, these fragments may represent a persistent component of the DART ejecta cloud and could be relevant to the observational targets of the upcoming Hera mission.

\vspace{0.5em}
\noindent{\sffamily\itshape -- Influence from cone-axis direction} \par
\vspace{0.3em}

The cone-axis direction has a much stronger influence on the fate statistics than the ejecta size. For the ``updated'' cone-axis direction, the escape fraction remains extremely high, generally above 99.4$\%$, while the impact fractions on both Didymos and Dimorphos remain below 0.6$\%$ for all size groups. In contrast, for the ``anti-DART'' cone-axis direction, the impact probability on Didymos increases substantially, reaching approximately 3.0\%--6.5\%, which depends on the size group and half-cone angle range. This suggests that the anti-DART-impact cone geometry preferentially directs a larger fraction of particles onto trajectories that re-impact Didymos. A detailed analysis of the initial directions associated with this specific fraction of particles is provided in the following section.

\vspace{0.5em}
\noindent{\sffamily\itshape -- Influence from half-cone angle} \par
\vspace{0.3em}

The half-cone angle range also affects the fate distribution, but its influence is secondary compared with the cone-axis direction. For both ``updated'' and ``anti-DART'' directions, changing the half-cone angle interval from [68$^\circ$, 72$^\circ$] to [62$^\circ$, 72$^\circ$] modifies the impact and escape fractions, with the most noticeable changes occurring for the ``anti-DART'' cases. In particular, the half-cone angle range [68$^\circ$, 72$^\circ$] with ``anti-DART'' cone-axis direction consistently produces higher Didymos impact fractions than the wider [62$^\circ$, 72$^\circ$] cone. This indicates that the narrower cone centred at larger half-cone angles is more favourable for generating Didymos-impacting trajectories. The influence of the cone-angle range is analysed in conjunction with the effect of the cone-axis direction and is discussed in the following context. 

In summary, Table~\ref{tab:EjectaFateStatistics} demonstrates that escape from the binary system is the dominant outcome for DART-generated ejecta particles, with more than 93.5\% escaping from the system within two-year time span. Nevertheless, a minute but dynamically relevant fraction of fragments remains in the near-binary environment over extended durations. Among the factors considered in Table~\ref{tab:EjectaFateStatistics}, the ejecta fate is primarily governed by the ejecta-cone geometry, especially the cone-axis direction, while particle size and cone-angle range provide secondary modulation. The simulations indicate that while most ejecta particles escape from the binary system, a non-negligible fraction can impact onto Didymos. The Didymos impact percentage increases further in the ``anti-DART'' cone-axis cases. This highlights the importance of accurately constraining the ejecta-cone geometry when modelling the long-term evolution and distribution of DART-generated ejecta in the Didymos system.

\section{Surface accretion of DART ejecta}
{\label{C5:Section5}}
This section analyses the surface accretion of DART impact ejecta onto Didymos and Dimorphos. Two complementary maps are constructed: the count-based impact density map and size-distribution-weighted surface accretion density map. The former describes the dynamical impact pattern of simulated ejecta particles, while the latter represents the estimated deposited ejecta mass per unit surface area. In addition, the influence of cone-axis direction and half-cone angle range on Didymos impacts is investigated through count-based impact density map.


\subsection{Count-based impact density maps}
\begin{figure}[ht!]
    \centering
    \includegraphics[width=0.9\linewidth]{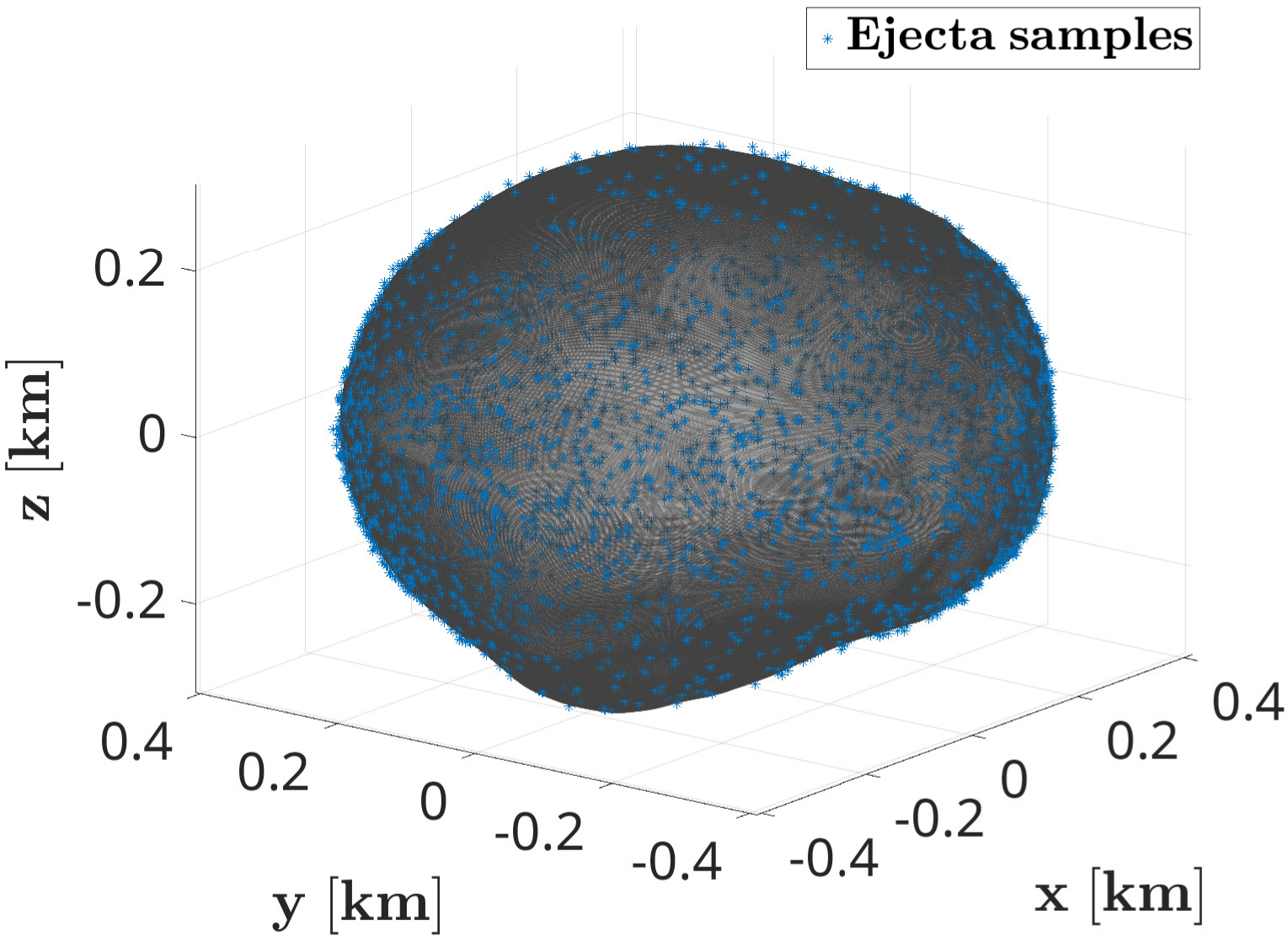}
    \caption{Illustration of ejecta samples impacting Didymos. The case shown corresponds to the ejecta size group $r_\text{ej}\in[10^{-2}, 10^{-1})$ m and ejecta cone geometry with ``updated'' cone-axis direction and half-cone angle range $(\theta_{\min}, \theta_{\max})=(68^\circ, 72^\circ)$.}
    \label{fig:ImpactDemo_Updated_HalfAngle6872}
\end{figure}

{\label{C51:SubSection51}}
Based on the ejecta evolution simulation results, count-based impact density maps are constructed directly from the recorded impact events. As illustrated in Fig.~\ref{fig:ImpactDemo_Updated_HalfAngle6872}, the adopted integration and event-detection scheme allows the ejecta impact locations on the polyhedral asteroid surface to be recorded accurately. By discretising the asteroid surface into 1$^\circ$ by 1$^\circ$ longitude-latitude bins, each impacting particle is assigned to the corresponding bin according to its impact longitude and latitude. The number of impacts in each bin is subsequently normalised by the bin's associated surface area to obtain the count-based impact density. 
\begin{figure*}[ht!]
    \centering
    \includegraphics[width=0.492\hsize]{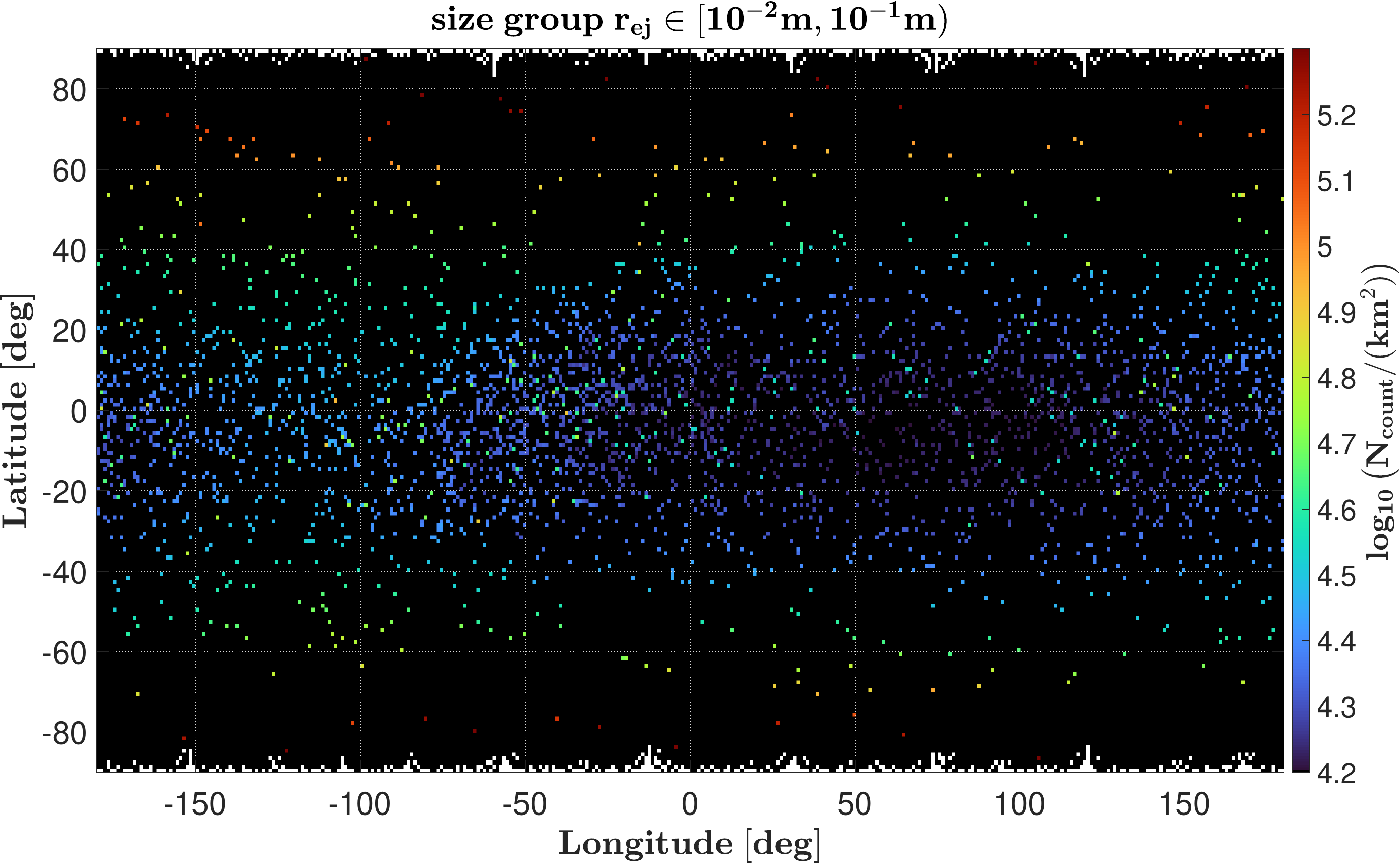} \;
    \includegraphics[width=0.492\hsize]{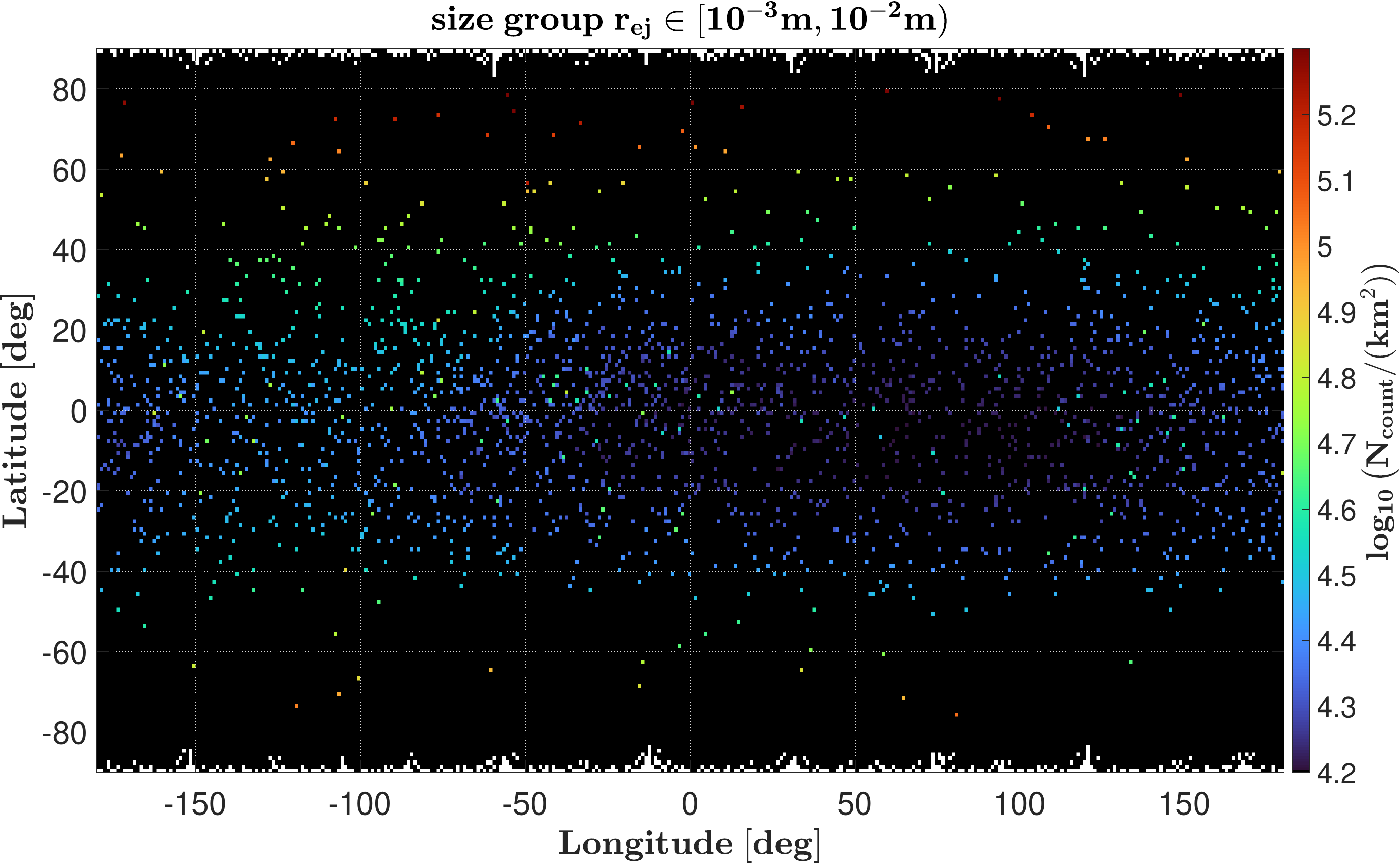}
    \includegraphics[width=0.492\hsize]{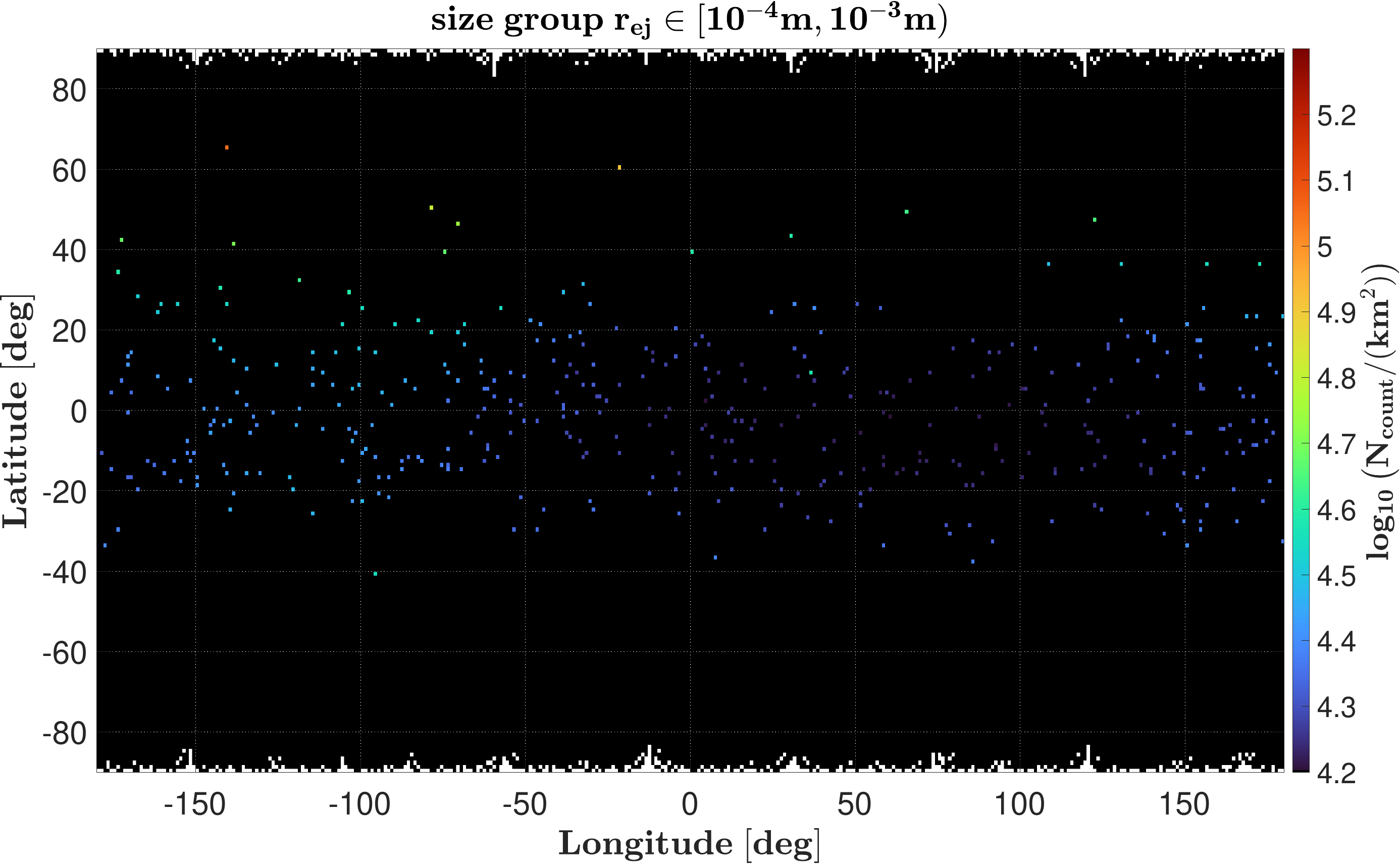} \;
    \includegraphics[width=0.492\hsize]{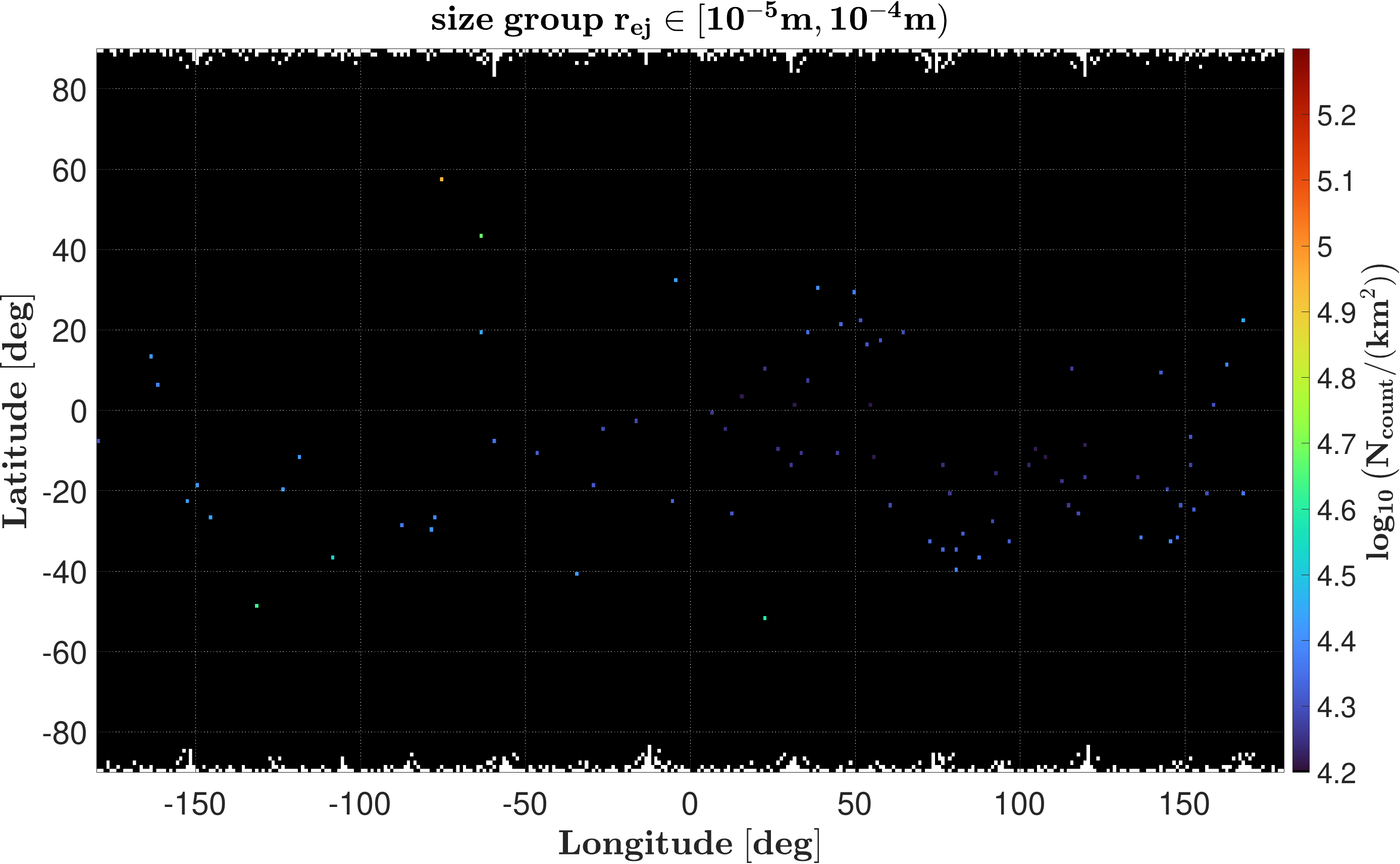}
    \includegraphics[width=0.492\hsize]{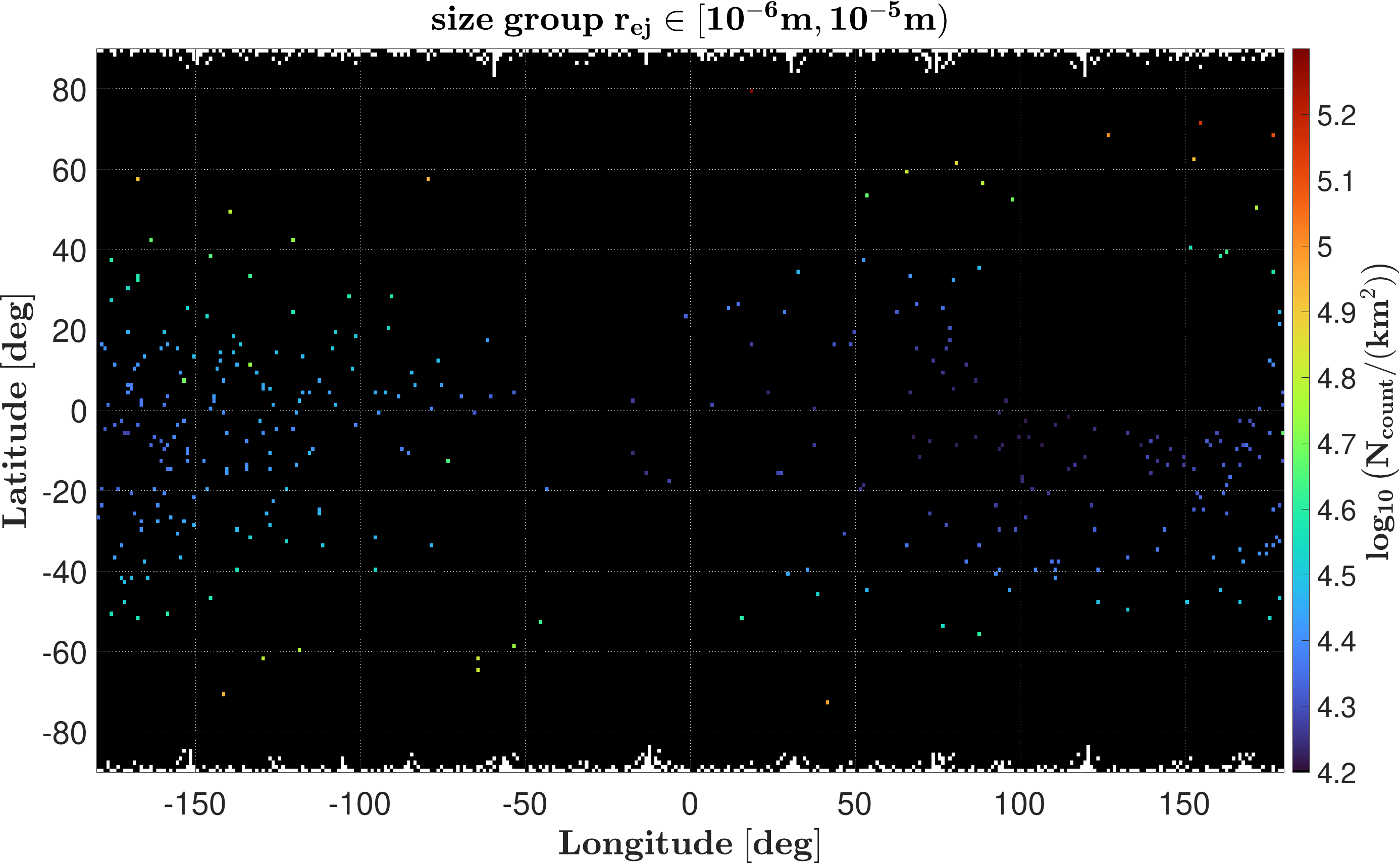}
    \caption{Count-based impact density maps of Didymos for the half-cone angle range $(\theta_{\min}, \theta_{\max})=(68^\circ, 72^\circ)$ and the ``updated'' cone-axis direction. From top to bottom and from left to right, the five panels correspond to ejecta size groups $r_{\text{ej}}=[10^{-(i+1)}, 10^{-i})$~m with $(i$ = $1,...,5)$, respectively. The count-based impact density is shown on a logarithmic scale.}
    \label{fig:CountBasedImpactDensityMaps_UpdatedHalfAngle6872}
\end{figure*}

\begin{figure*}[ht!]
    \centering
    \includegraphics[width=0.492\hsize]{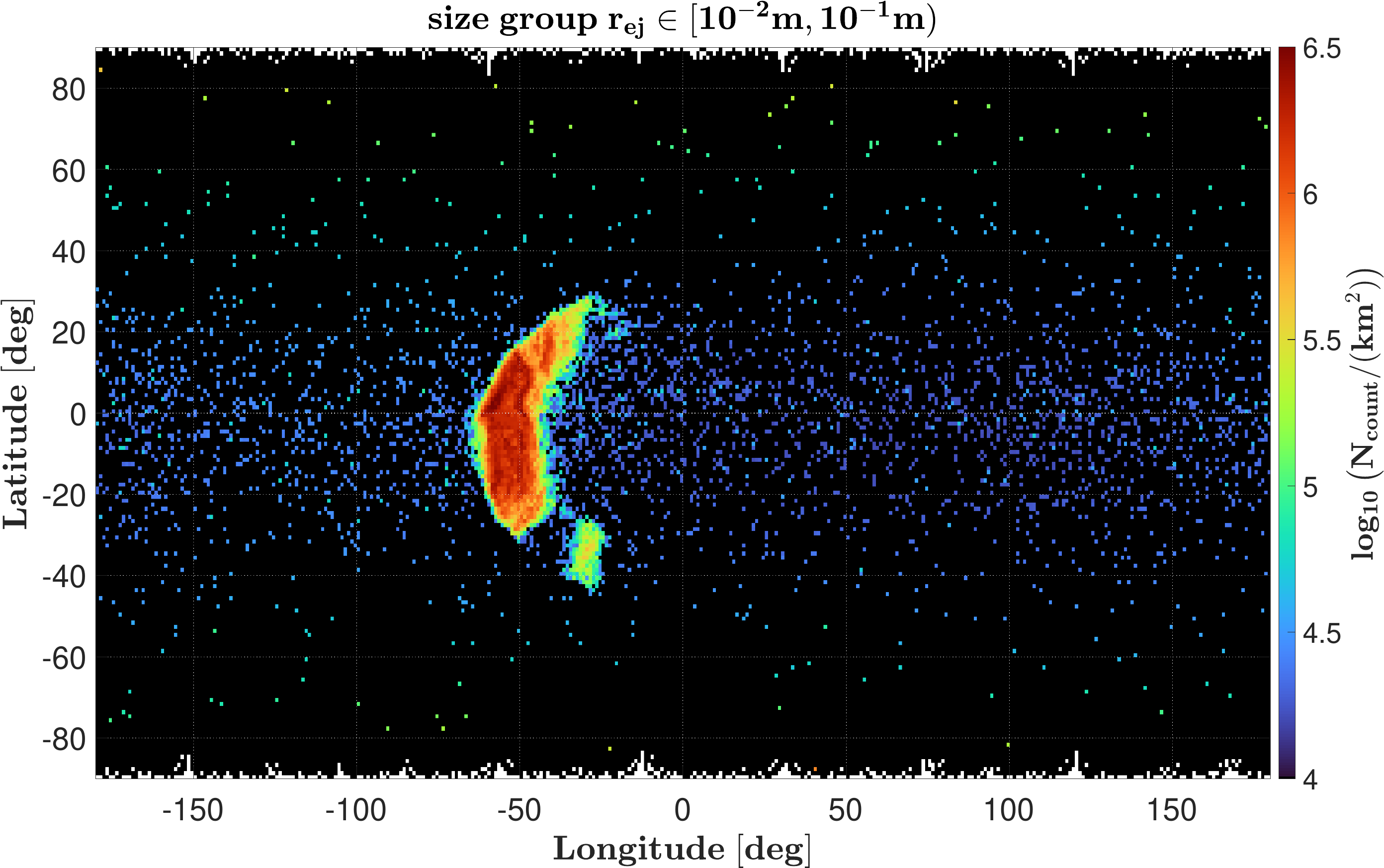} \;
    \includegraphics[width=0.492\hsize]{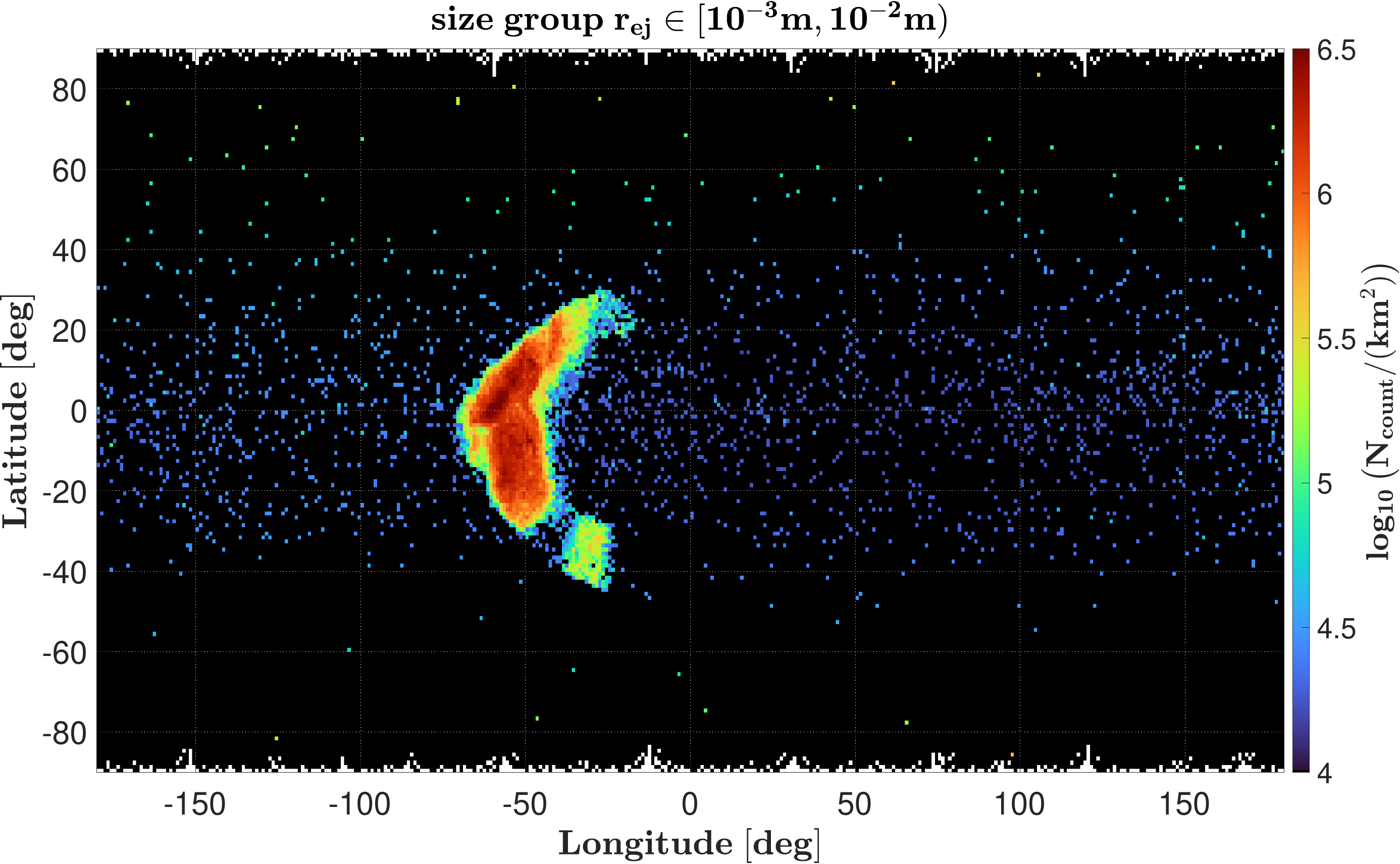}
    \includegraphics[width=0.492\hsize]{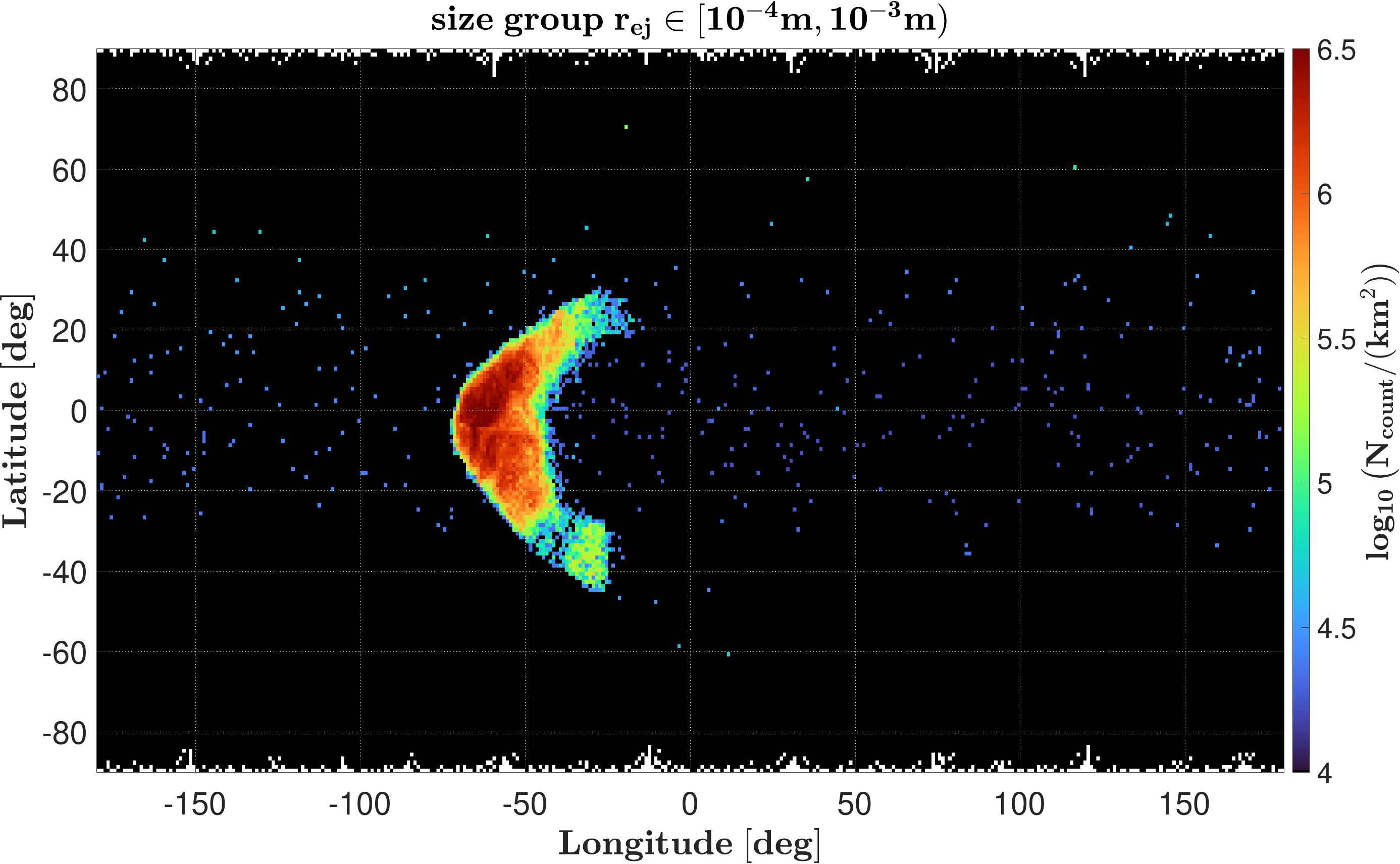} \;
    \includegraphics[width=0.492\hsize]{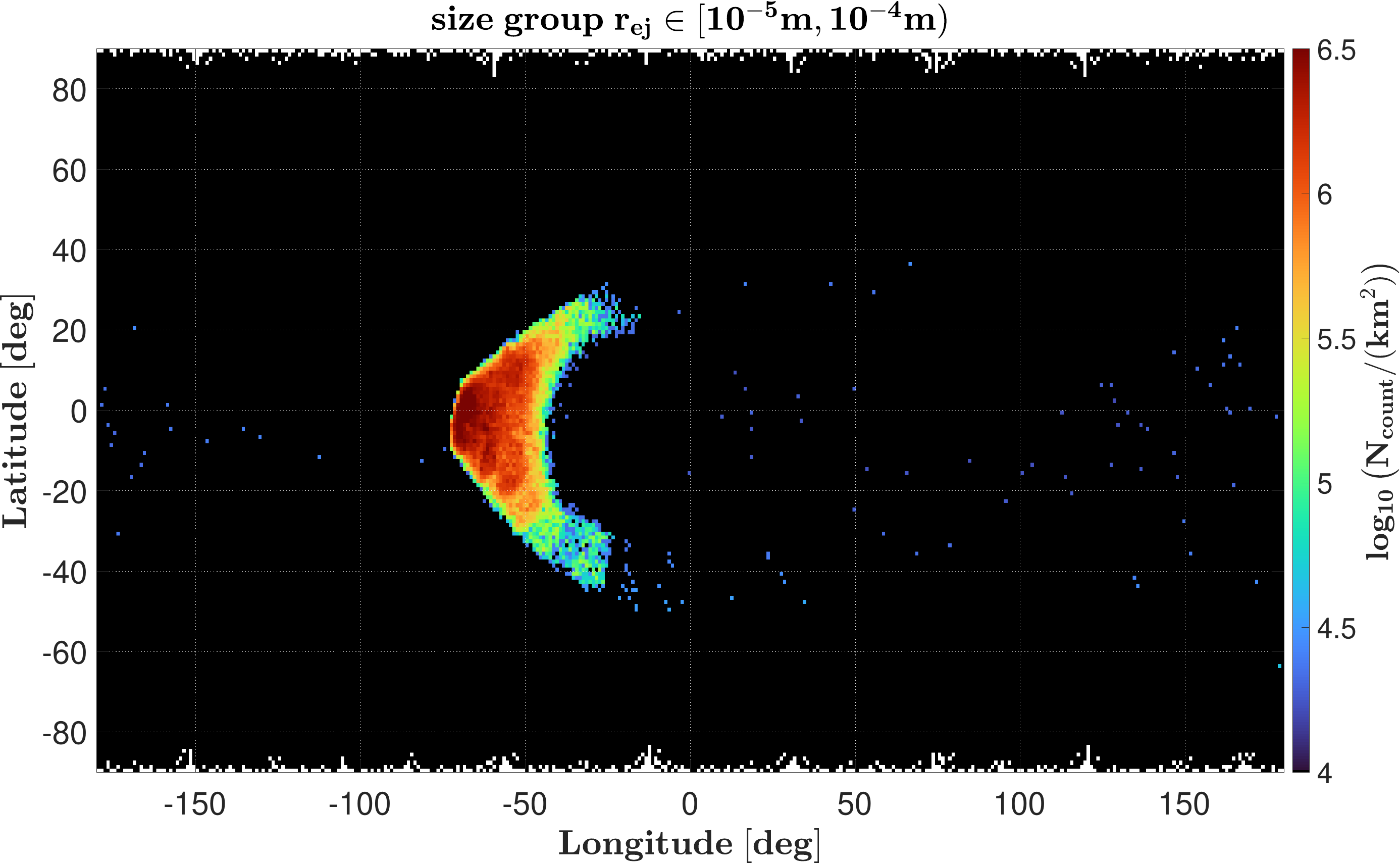}
    \includegraphics[width=0.492\hsize]{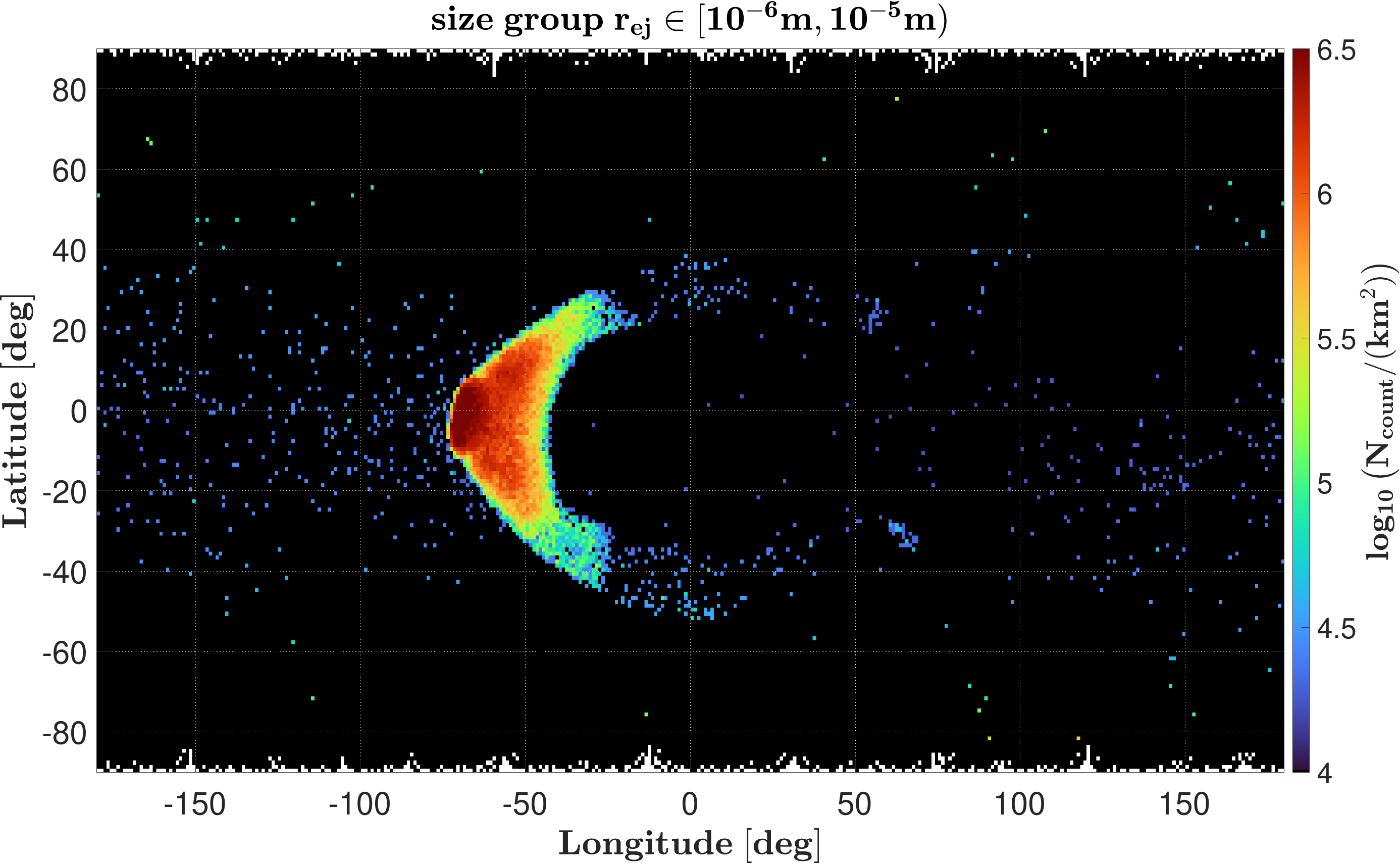}
    \caption{Count-based impact density maps of Didymos for the half-cone angle range $(\theta_{\min}, \theta_{\max})=(68^\circ, 72^\circ)$ and the anti-DART-impact cone-axis direction. From top to bottom and from left to right, the five panels correspond to ejecta size groups $r_{\text{ej}}=[10^{-(i+1)}, 10^{-i})$~m with $(i$ = $1,...,5)$, respectively. The count-based impact density is shown on a logarithmic scale.}
    \label{fig:CountBasedImpactDensityMaps_OppoDARTHalfAngle6872}
\end{figure*}

The count-based impact density maps of the half-cone angle range $(\theta_{\min}, \theta_{\max})=(68^\circ, 72^\circ)$ with the two candidate cone-axis directions are shown in Figs.~\ref{fig:CountBasedImpactDensityMaps_UpdatedHalfAngle6872} and~\ref{fig:CountBasedImpactDensityMaps_OppoDARTHalfAngle6872}, respectively. Note that in both figures, the empty blocks near the poles are caused by the limited resolution of the selected shape model in those regions. More specifically, the small number of triangular facets representing the polar topography does not guarantee that each longitude-latitude bin can be assigned a non-zero surface area, leading to a zero denominator when computing the impact density. However, this limitation does not affect the overall impact density distribution, since the simulations show that no more than 0.001\% of sampled particles reach the polar regions in any considered cases. The analyses and discussion of the two count-based impact density maps are presented separately below.

\vspace{0.5em}
\noindent{\sffamily\itshape -- Maps associated with ``updated'' cone-axis direction} \par
\vspace{0.3em}

As shown in Fig.~\ref{fig:CountBasedImpactDensityMaps_UpdatedHalfAngle6872}, for the half-cone angle range $(\theta_{\min}, \theta_{\max})=(68^\circ, 72^\circ)$ and the ``updated'' cone-axis direction, the ejecta impacting onto Didymos is not uniformly distributed over the surface. For all five size groups, the impacts are concentrated mainly at low-to-mid latitudes, with most populated bins located approximately between longitudes $-40^{\circ}$ and $40^{\circ}$. High-latitude impacts are comparatively sparse, and the polar regions contain almost no ejecta samples. Therefore, the dominant accretion pattern is a broad, longitudinally extended impact band rather than a globally uniform deposition pattern. 

Additionally, a clear ejecta size dependence is identified in Fig.~\ref{fig:CountBasedImpactDensityMaps_UpdatedHalfAngle6872}. The two largest size groups, $r_{\text{ej}}=[10^{-2}, 10^{-1})$~m and $[10^{-3}, 10^{-2})$~m, produce the densest and most spatially continuous impact distributions. Their impacts occupy a wide range of longitudes and latitudes, with locally enhanced count-based impact densities reaching the upper part of the plotted logarithmic scale. This indicates that, among the particles impacting Didymos, larger ejecta fragments are more likely to produce spatially extensive accretion patterns. In comparison, the rest three smaller size groups show more scattered and discontinuous impact patterns. The impacts of the intermediate size group, $r_{\text{ej}}=[10^{-4}, 10^{-3})$~m, are still mainly confined to low-to-mid latitudes, but the number of populated bins is much smaller than those of the two larger size groups. The $r_{\text{ej}}=[10^{-5}, 10^{-4})$~m group has very few populated bins, whereas the $r_{\text{ej}}=[10^{-6}, 10^{-5})$~m group shows a relatively broader but still sparse distribution. This non-monotonic behaviour is physically plausible because smaller particles are more sensitive to the SRP influence. Instead of continuing along gravity-dominated trajectories that lead to surface impact, these particles can experience stronger trajectory deviations under SRP. Consequently, a larger fraction of these particles is dispersed away from the impact-favourable regions, resulting in a reduced particle count and a much sparser impact-density distribution. Overall, Fig.~\ref{fig:CountBasedImpactDensityMaps_UpdatedHalfAngle6872} indicates that Didymos accretion is dominated by low-to-mid-latitude impacts, with the two largest ejecta size groups producing most spatially extensive and dense count-based impact distributions.

\begin{figure*}[ht!]
    \centering
    \includegraphics[width=0.32\hsize]{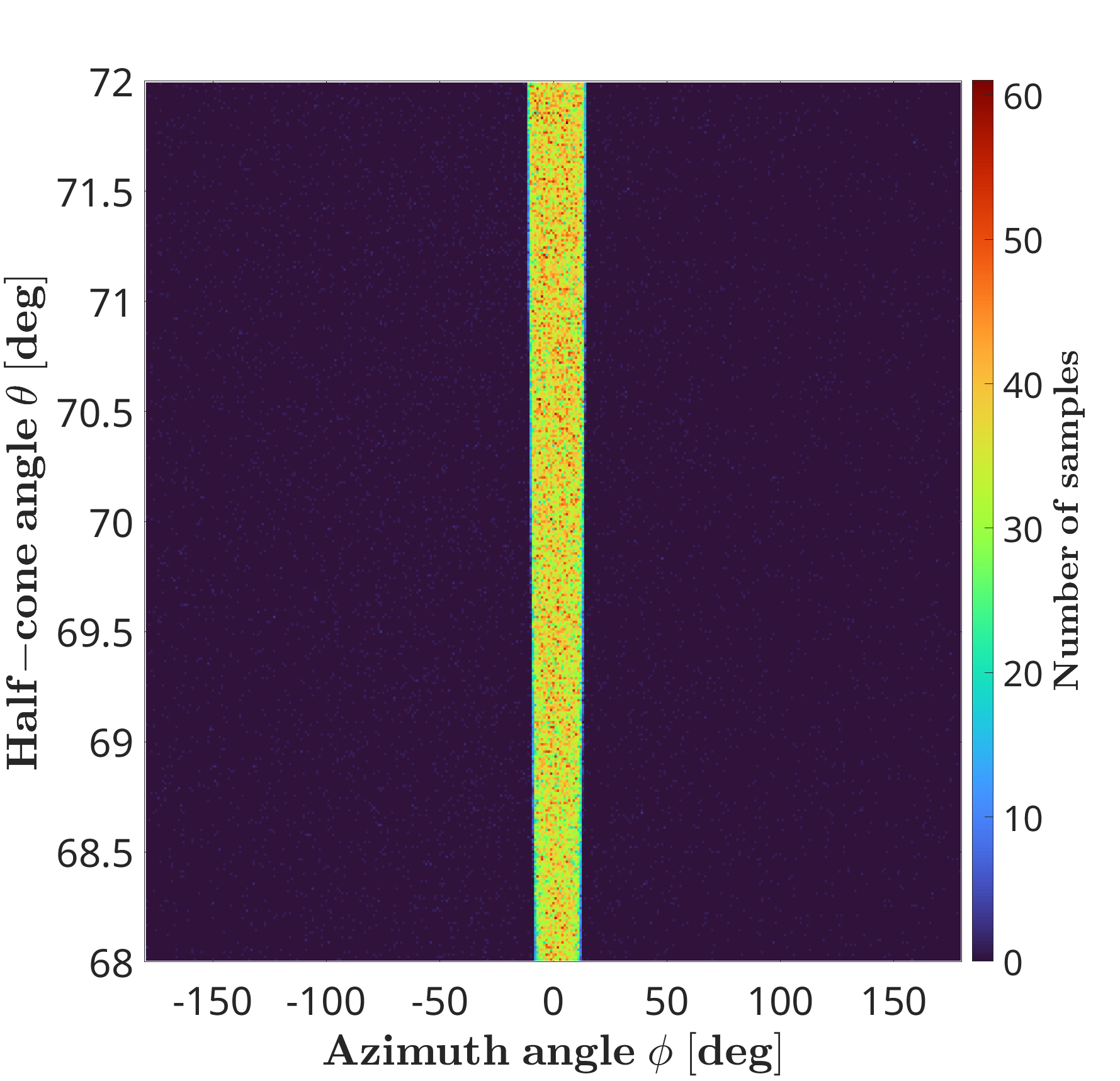}
    \includegraphics[width=0.3\hsize]{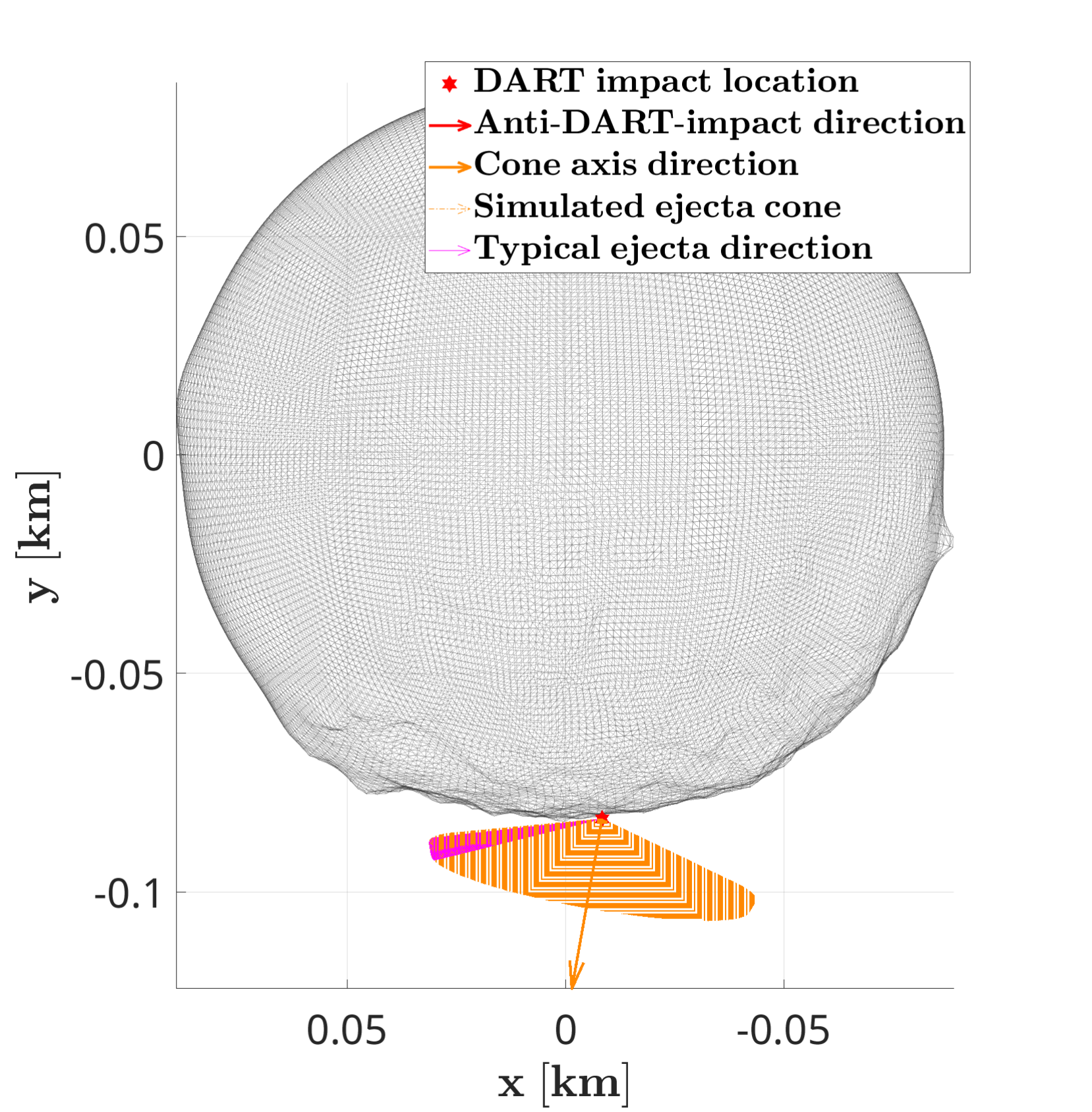}
    \includegraphics[width=0.323\hsize]{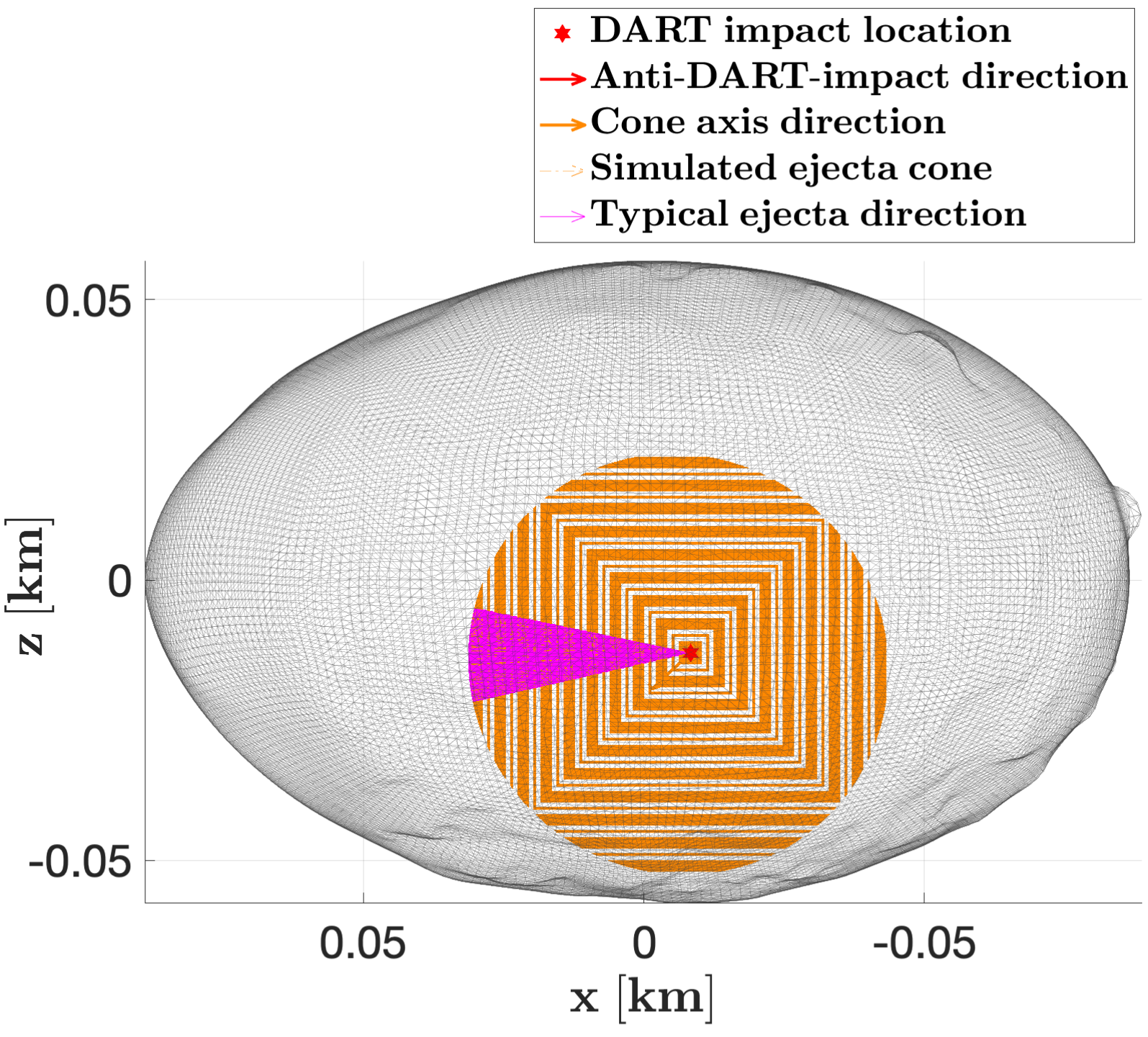}
    \caption{Illustration of the ejecta sample subset associated with localised high impact-density feature. The left panel shows the azimuth and half-cone-angle distribution of this subset in the cone-centred coordinates. The middle and right panels show the corresponding initial velocity directions, indicated by magenta arrows, viewed from the south-pole direction (same as the third panel in Fig.~\ref{fig:EjectaConeGeometries}) and from 270$^{\circ}$E, respectively.}
    \label{fig:TypicalDirections}
\end{figure*}

\vspace{0.5em}
\noindent{\sffamily\itshape -- Maps associated with ``anti-DART'' cone-axis direction} \par
\vspace{0.3em}

In contrast, as shown in Fig.~\ref{fig:CountBasedImpactDensityMaps_OppoDARTHalfAngle6872}, the count-based impact density map for the half-cone angle range $(\theta_{\min}, \theta_{\max})=(68^\circ, 72^\circ)$ and the ``anti-DART'' cone-axis direction exhibits a more localised accretion pattern on Didymos than the ``updated'' cone-axis case. For all five ejecta size groups, the dominant impacts are concentrated in a well-defined region centred approximately between longitudes -$70^{\circ}$ and -$20^{\circ}$, with most high-density bins lying between latitudes about -$30^{\circ}$ and -$30^{\circ}$. On the logarithmic colour scale, the densest bins in this region reach values close to $\text{log}_{10}(N_\text{count}/\text{km}^2) \approx$ 6--6.5, indicating that a subset of impact samples is geometrically focused onto a relatively narrow surface region on Didymos.

Guided by this localised count-based impact-density feature, the corresponding subset of impact samples is identified within the ejecta cone. As shown in the left panel of Fig.~\ref{fig:TypicalDirections}, these samples are concentrated within azimuth angles of $\phi\in (-12.5 ^{\circ},12.5 ^{\circ})$ in the cone-centred coordinates. Their associated initial velocity directions are also illustrated in Fig.~\ref{fig:TypicalDirections} to provide a clearer geometrical interpretation. Apart from this angular concentration, these samples also exhibit a typical speed distribution. Among the 299,601 samples reaching Dimorphos's surface for this specific cone geometry, 291,002 samples are located within the condensed high-density region. Within this region, 95.71\% of the samples have initial speeds greater than 3~m/s, and 86.26\% have initial speed greater than 6~m/s. These relatively high initial speeds enables the particles to reach the surface of Didymos within no more than 5 minutes. 

These results suggest that the surface accretion pattern is governed by the coupled effects of the initial ejecta-cone geometry, the instantaneous configuration of the binary system at impact epoch, and the subsequent dynamical evolution of ejecta particles. A geometrical interpretation can be obtained by comparing the impact-sample distribution shown in Fig.~\ref{fig:TypicalDirections} with the ejecta-cone geometries illustrated in the first panel of Fig.~\ref{fig:EjectaConeGeometries}. The Didymos-impacting samples are concentrated within a relatively narrow region of the cone-centred angular space as shown in Fig.~\ref{fig:TypicalDirections}. Therefore, when the ``anti-DART'' cone-axis is shifted towards the ``updated'' cone-axis direction, a portion of the high-speed samples that would otherwise be directed toward Didymos is redistributed away from the impact-favourable angular region. These particles consequently avoid direct encounter with Didymos and evolve towards different final fates.

A similar interpretation applies when the ``anti-DART'' cone-axis direction is fixed but the half-cone-angle range is widened to $[62^{\circ},72 ^{\circ}]$. In this case, the broader angular sampling redirects part of the originally impact-favourable population into additional ejection directions that do not lead to Didymos impact. As a result, the narrower $[68^{\circ},72 ^{\circ}]$ cone retains a larger fraction of particles within the Didymos-impact-favourable angular region, whereas the wider cone distributes the samples over a larger solid-angle domain and reduces the relative contribution of direct-impact trajectories. This geometrical interpretation explains the difference in the Didymos-impact fractions between the wide and narrow half-cone angle cases reported in Table~\ref{tab:EjectaFateStatistics}. 

The difference is also evident in the corresponding impact density maps for the two half-cone angle ranges. Although the maps associated with the wider half-cone angle range are still characterised by the localised crescent-shaped distribution pattern in the same region, the impact density in that region is lower than that obtained for the narrower cone. Therefore, for conciseness, only the maps for the half-cone angle range $(\theta_{\min}, \theta_{\max})=(68^\circ, 72^\circ)$ are included in this section. 

It is remarked that, at the same time, after removing this compact crescent-shaped high-density structure, a broader longitudinally extended impact pattern, similar to the one observed in the ``updated'' cone-axis case shown in Fig.~\ref{fig:CountBasedImpactDensityMaps_UpdatedHalfAngle6872}, also emerges. This further indicates that the impact-density distribution induced by DART impact may consist of both a strongly focused early accretion component and a more diffuse background accretion component. 

Such a finding may also be relevant to the hypothesised secondary ejection event caused by debris impacting Didymos, as discussed by \cite{Moreno2023characterization}. In their study, the northern component of the observed double tail, first seen from 2022 October 8, was suggested as a possible signature of secondary ejection triggered by ejecta impacting Didymos. The present simulations provide a dynamical pathway consistent with this interpretation: a concentrated subset of particles is found to re-impact Didymos, particularly for ejection geometries close to the ``anti-DART'' direction. Since this Didymos-impacting population is identified across different ejecta size groups, the mechanism is not restricted to a single particle-size range within the present sampling. It is therefore plausible that larger fragments or boulders ejected along similar impact-favourable directions could produce higher-energy sesquinary impacts on Didymos, potentially releasing secondary ejecta that contributes to the observed double-tail morphology. Nevertheless, this interpretation remains qualitative, as the present study does not explicitly model the impact-generated secondary ejecta or its subsequent optical observability.

In addition to the maps generated for Didymos, the count-based impact density maps for Dimorphos are also produced. They exhibit a broadly similar longitudinally extended impact pattern, with relatively higher densities at mid latitudes while lower densities at low latitudes. Same as the Didymos case, the two largest size groups, $r_{\text{ej}}=[10^{-2}, 10^{-1})$~m and $[10^{-3}, 10^{-2})$~m, generate the densest and most spatially continuous impact distributions. Their impacts occupy a wide range of longitudes and latitudes. In contrast, the contributions from the three smaller size groups are comparatively limited. In addition, no localised high-density region, comparable to the compact accretion feature identified on Didymos, is observed for Dimorphos under the ``anti-DART'' cone-axis direction. Therefore, for conciseness, the corresponding count-based impact density maps of Dimorphos are not shown here. The more physically informative size-distribution-weighted surface accretion maps for Dimorphos are instead presented in the following subsection.

\subsection{Surface accretion density maps}
{\label{C52:SubSection52}}

The count-based impact-density maps describe the spatial distribution of the simulated particles for different ejecta size groups. However, they do not directly represent the physical ejecta mass deposited on the asteroid surface, since the real ejecta population is expected to follow a size-frequency distribution. To estimate the physical surface accretion pattern, the simulated impact statistics must be further weighted by an ejecta size-frequency distribution. In this study, the physical mass contribution of each ejecta size group is estimated using the broken power-law size distribution derived from the HST observations for DART impact \citep{Kim2023Single}. Let \(p(r_\text{ej})\) denote the differential number distribution of ejecta particle radius $r_\text{ej}$, the adopted size-frequency distribution is written as:
\begin{equation}
p(r_\text{ej})
=
\begin{cases}
P_1\,r_\text{ej}^{-2.7}, & 1\times10^{-6}\le r_\text{ej} \le 2\times10^{-3}~\mathrm{m}\\
P_2\,r_\text{ej}^{-3.9}, & 2\times10^{-3}< r_\text{ej} \le 1\times10^{-2}~\mathrm{m}\\
P_3\,r_\text{ej}^{-4.2}, & 1\times10^{-2}< r_\text{ej} \le 2\times10^{-1}~\mathrm{m}
\end{cases}
\label{eq:SizeFrequencyDistribution}
\end{equation}
where \(P_1\), \(P_2\), and \(P_3\) are the reference particle production rates in the three radius ranges, and their values are derived as: \(P_2=480\), $P_1=P_2\times0.002^{(2.7-3.9)}=8.3177\times10^{5}$, and $P_3=P_2\times0.01^{(4.2-3.9)}=120.5705$. 

\begin{figure*}[ht!]
    \centering
    \includegraphics[width=0.493\hsize]{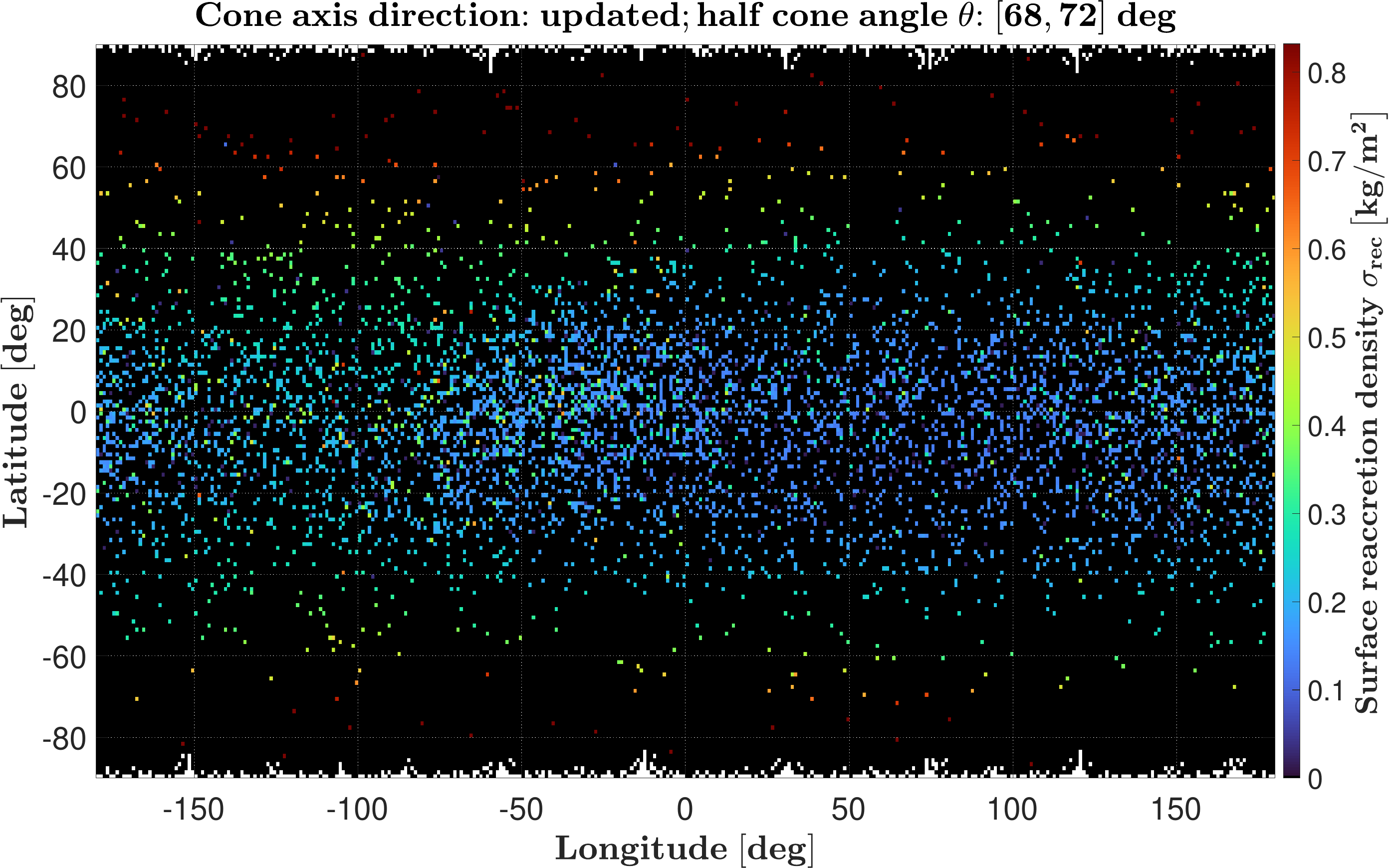} \;
    \includegraphics[width=0.49\hsize]{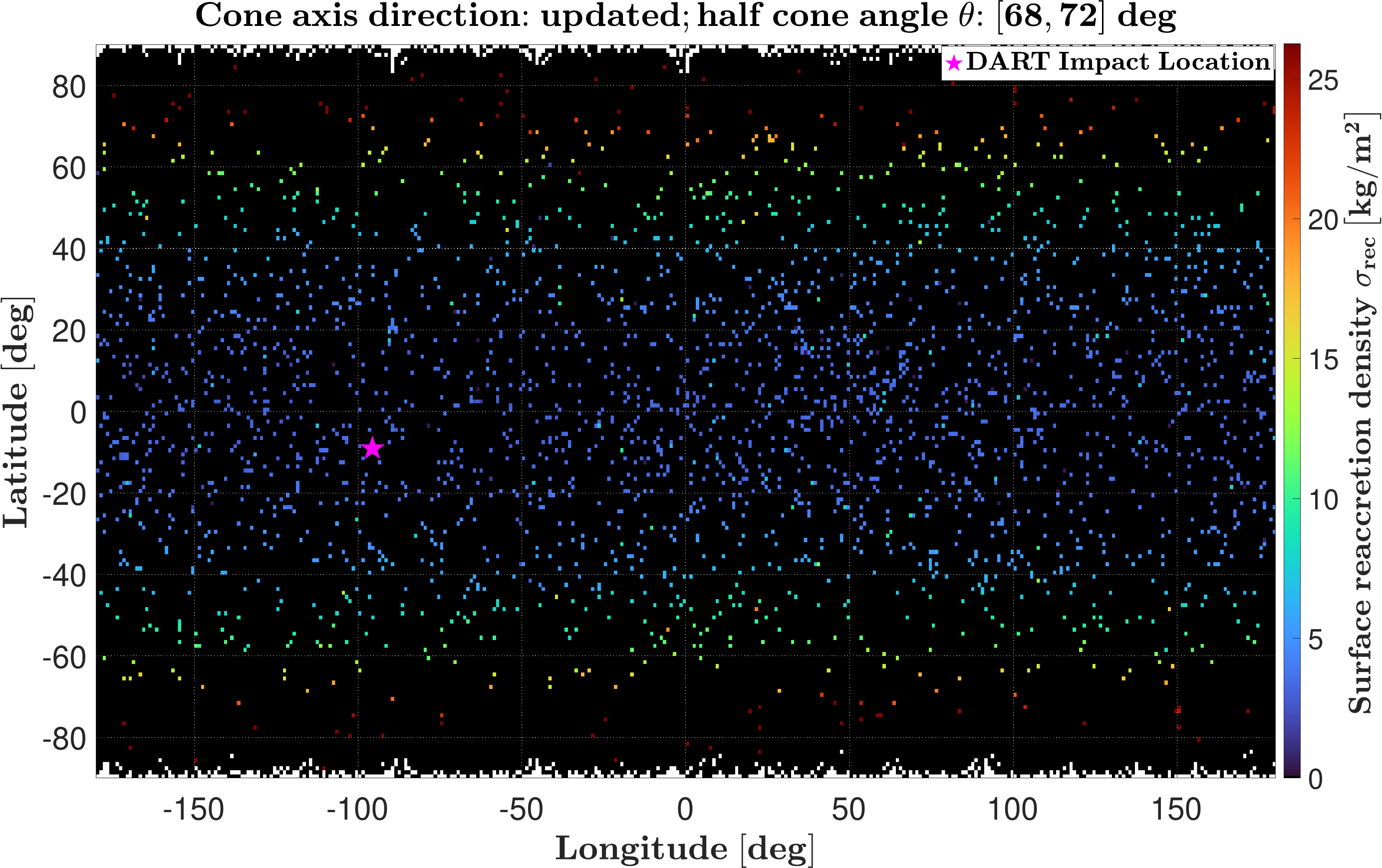}
    \includegraphics[width=0.493\hsize]{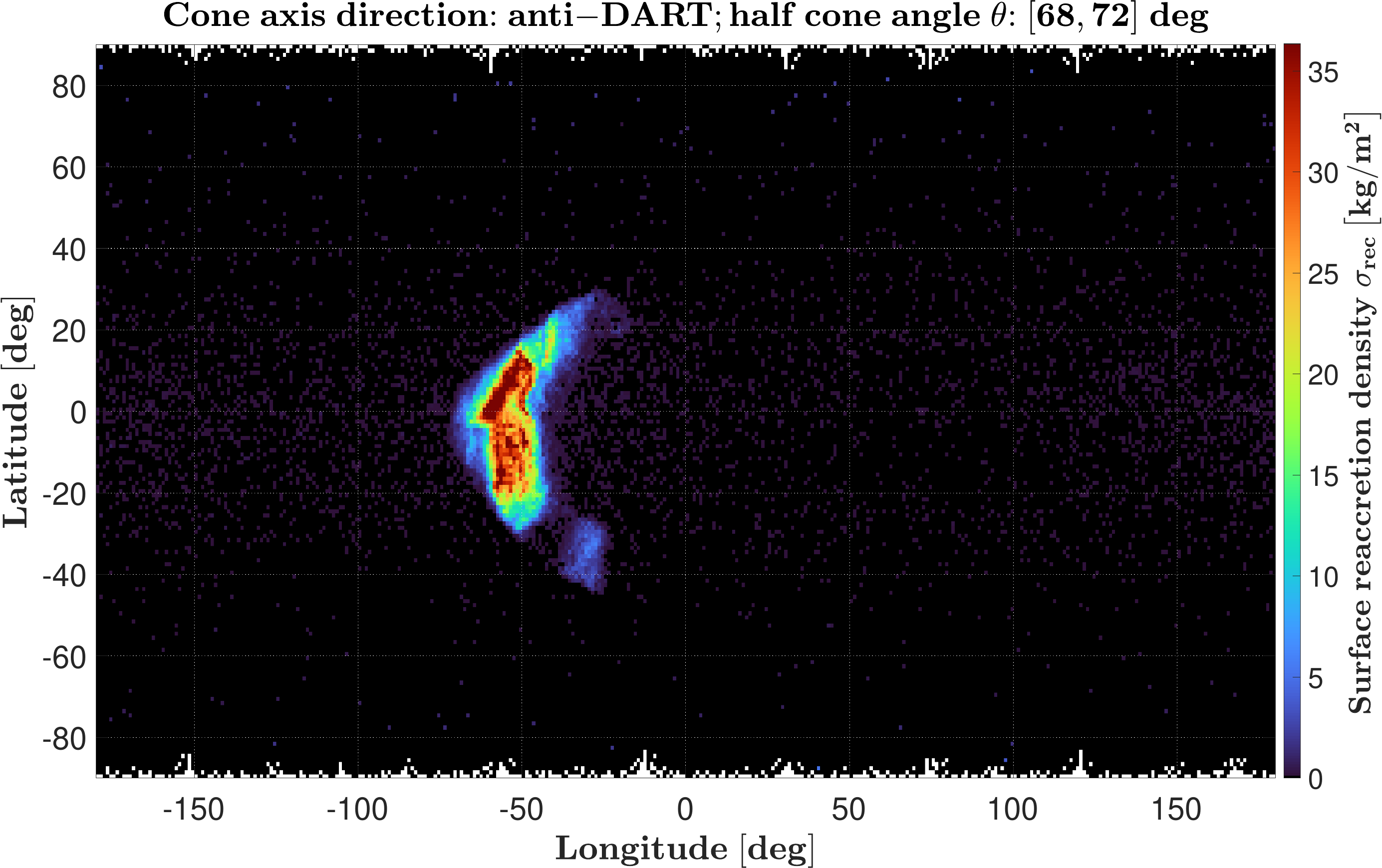} \;
    \includegraphics[width=0.49\hsize]{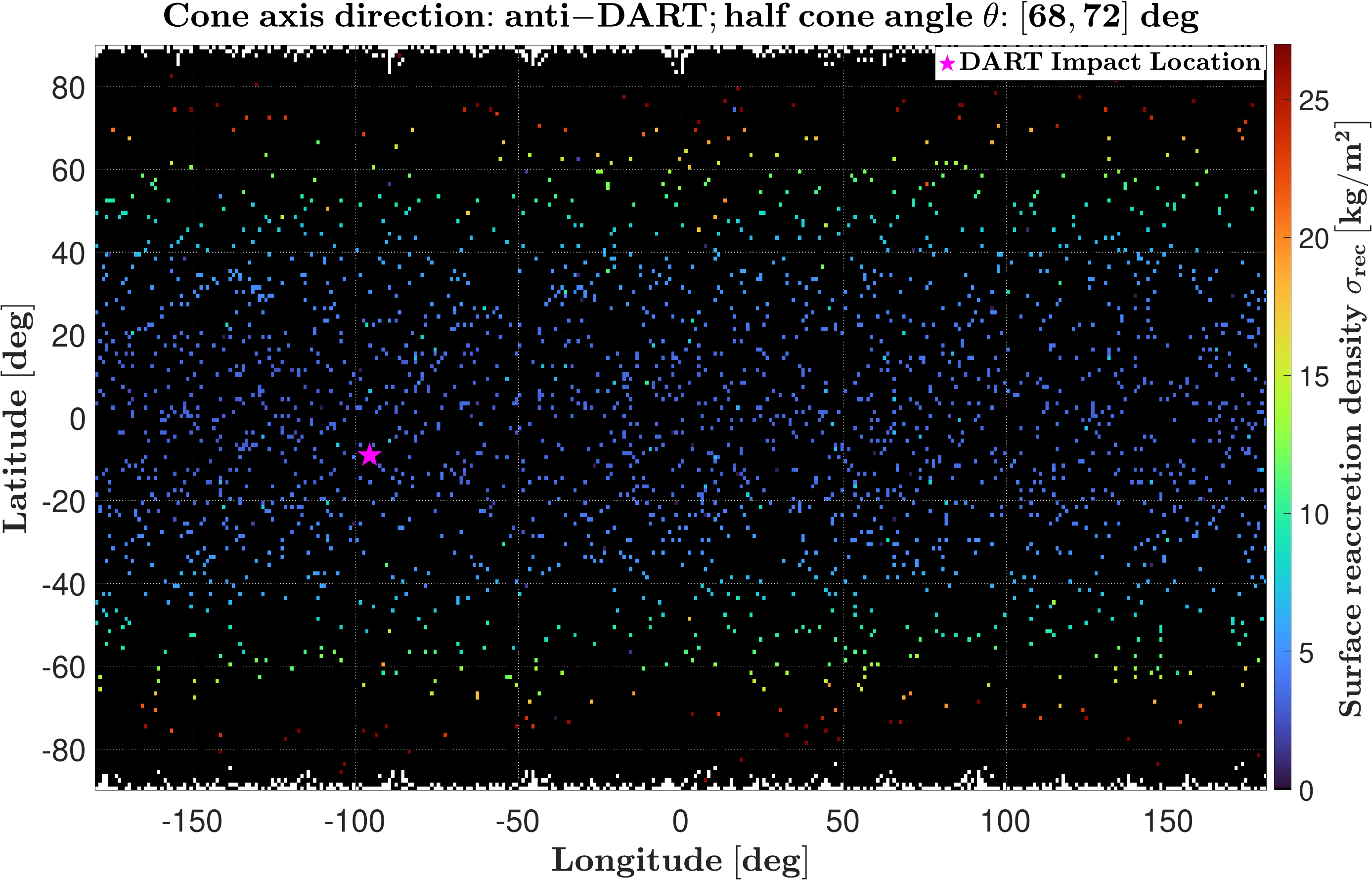}
    \caption{Surface accretion density maps of Didymos and Dimorphos for two candidate ejecta cone geometries with the half-cone angle range $(\theta_{\min}, \theta_{\max})=(68^\circ, 72^\circ)$. The left two panels correspond to Didymos, and the right two panels correspond to Dimorphos. The top two panels are associated with the ``updated'' cone-axis direction, whereas the bottom two panels are associated with the ``anti-DART'' cone-axis direction.}
    \label{fig:SurfaceReaccretionDensity2DMap}
\end{figure*}

Based on the adopted size-frequency distribution, the procedure for estimating the size-distribution weighted surface accretion density is as follows. Let the differential size-frequency distribution in Eq.~(\ref{eq:SizeFrequencyDistribution}) be simplified in the compact form $p(r_\text{ej}) = P(r)\,r^{-q(r)}$, where $P(r)$ denotes the piecewise particle production rate and $q(r)$ denotes the corresponding power-law index. Assuming spherical particles with constant bulk density, the mass fraction associated with the \(i\)-th size group is computed by integrating the mass-weighted size distribution over the corresponding radius interval:
\begin{equation}
f_i =
\frac{
\displaystyle \int_{r_{i,\min}}^{r_{i,\max}}P(r)\,r^{3-q(r)}\,\mathrm{d}r
}{
\displaystyle \sum_{k=1}^{5}
\int_{r_{k,\min}}^{r_{k,\max}}P(r)\,r^{3-q(r)}\,\mathrm{d}r
}
\label{eq:MassFraction}
\end{equation}
where \(r_{i,\min}\) and \(r_{i,\max}\) are the lower and upper radius bounds of the \(i\)-th size group. Denoting the total ejecta mass by \(m_{\mathrm{total}}\), the mass accreted in a given longitude--latitude surface bin \(j\), $m_j^\text{rec}$, is estimated as:
\begin{equation}
m_j^\text{rec} 
=
m_{\mathrm{total}}
\sum_{i=1}^{5}
f_i \frac{N_{ij}}{N_i}
\end{equation}
where \(N_{ij}\) is the number of simulated particles from size group \(i\) that impact bin \(j\), and \(N_i\) is the total number of simulated particles in that size group. Finally, the size-distribution-weighted surface accretion density $\sigma_j^{\mathrm{rec}}$ is obtained by normalising this deposited mass by the surface area \(A_j\) of the corresponding bin:
\begin{equation}
\sigma_j^{\mathrm{rec}}
=
\frac{m_j^\text{rec}}{A_j}
=
\frac{m_{\mathrm{total}}}{A_j}
\sum_{i=1}^{5}
f_i \frac{N_{ij}}{N_i}
\end{equation}
The resulting quantity represents the estimated ejecta mass deposited per unit surface area. It is noteworthy that $\sigma_{\mathrm{rec}}$ is a model-dependent estimate which is influenced by the adopted ejecta size-frequency distribution, total ejecta mass, and the assumed representativeness of the simulated samples within each size group.

Using the reference value of total mass $1.7\times10^7$ kg reported by \cite{Kim2023Single} for ejecta particle size range from 1 $\mu m$ to 0.2 m, the surface accretion density maps of Didymos and Dimorphos are generated for the two candidate ejecta cone geometries with the half-cone angle range $(\theta_{\min}, \theta_{\max})=(68^\circ, 72^\circ)$. As shown in Fig.~\ref{fig:SurfaceReaccretionDensity2DMap}, the left column shows the accretion pattern on Didymos, whereas the right column shows the corresponding pattern on Dimorphos. The top row corresponds to the ``updated'' cone-axis direction, and the bottom row corresponds to the ``anti-DART'' cone-axis direction.

For Didymos, the two cone-axis directions produce substantially different surface accretion patterns. In the ``updated'' cone-axis case, the re-accreted ejecta is distributed over a broad range of longitudes and is mainly concentrated at low-to-mid latitudes, approximately within -$40^{\circ}$ to $40^{\circ}$. The surface accretion density is relatively diffuse, with values mostly below 1 kg$/\text{m}^2$. This indicates that, although ejecta particles impact Didymos over a large surface area, the deposited mass per unit area remains modest. In contrast, the ``anti-DART'' case produces a much more localised and intense accretion footprint on Didymos. The highest-density region forms a compact crescent-shaped structure centred roughly between longitudes \(-70^\circ\) and \(-30^\circ\), mainly within latitudes \(-30^\circ\) to \(25^\circ\). The peak surface accretion density reaches around 40 kg$/\text{m}^2$, substantially higher than in the ``updated'' case. The difference identified in the left two panels in Fig.~\ref{fig:SurfaceReaccretionDensity2DMap} is consistent with the comparison made among the count-based impact density maps in Figs.~\ref{fig:CountBasedImpactDensityMaps_UpdatedHalfAngle6872} and \ref{fig:CountBasedImpactDensityMaps_OppoDARTHalfAngle6872}. This demonstrates that the cone-axis direction strongly controls not only the number of impact particles, but also the spatial concentration of deposited ejecta mass on Didymos.

For Dimorphos, the accretion density maps are more broadly distributed and less strongly localised than the high-density Didymos footprint in the ``anti-DART'' case. In both cone-axis configurations, the deposited material spans a wide range of longitudes and latitudes, with the DART impact location marked by the magenta star. The accretion density tends to be higher at high northern and southern latitudes than near the equatorial region, although the distribution remains spatially scattered. The colour scales indicate that local surface accretion densities on Dimorphos can reach values of order \textbf{\(5\)--\(25~\mathrm{kg\,m^{-2}}\)}, but without forming the same compact crescent-shaped concentration seen on Didymos for the ``anti-DART'' geometry. It should be noted that the accretion density maps for Dimorphos are generated using the latest available pre-impact shape model of Dimorphos. Consequently, the associated spatial distribution of surface impacts is referenced to the pre-impact surface geometry. The interpretation of these maps remains valid as long as the DART impact did not induce substantial global deformation and chaotic tumbling status of Dimorphos, although local morphological changes near the impact site may introduce additional uncertainty.

In summary, Fig.~\ref{fig:SurfaceReaccretionDensity2DMap} shows that surface accretion is highly sensitive to the adopted ejecta-cone geometries. The ``updated'' cone-axis direction leads to relatively diffuse deposition on Didymos, with $\sigma_\text{rec}$ mostly of the order of 0.3 kg$/\text{m}^2$ at low-to-mid latitudes. In contrast, the ``anti-DART'' cone-axis direction produces a concentrated crescent-shaped high-density signature on the primary, where $\sigma_\text{rec}$ increases to values exceeding 10 kg$/\text{m}^2$. Dimorphos exhibits broader and more dispersed accretion patterns for both cone-axis directions. The accretion density tends to be higher at high latitudes than mid-to-low latitudes, reaching values of order 5--25~kg$/\text{m}^2$ overall.

Based on the surface accretion density maps obtained for both asteroids, the equivalent thickness of the accreted ejecta layer is further estimated. Assuming a total mass $m_\text{total}$ of $1.7\times10^7$ kg involved in the ejecta-trail formation and a particle density of 3500~kg$/$m$^3$, as reported by \cite{Kim2023Single} and \cite{Ferrari2025Morphology}, the deposited layer thickness on Dimorphos is estimated to be of the order of 1.5 mm at mid-to-low latitudes. This estimate assumes complete horizontal spreading of the deposited material within each surface bin and zero porosity in the accreted layer. It also neglects re-ejection and any deformation or interaction between the impacting ejecta particles and the surface. For Didymos, in the case associated with the ``updated'' cone-axis direction, the re-accreted layer remains thinner than approximately 0.3 mm over the surface. In contrast, if the ``anti-DART'' cone-axis direction is adopted, the crescent-shaped high-density region on Didymos corresponds to an equivalent accreted layer thickness of approximately 3--11.5 mm. These estimates indicate that the predicted spatial distribution and thickness of accreted ejecta may serve as a useful reference for interpreting future in-situ observations by the Hera spacecraft.

\section{Conclusions}
{\label{C6:Section6}}
This study investigated the dynamical evolution of DART-generated impact ejecta and quantified their surface accretion patterns within the Didymos binary system. A high-fidelity dynamical model is constructed, incorporating polyhedron gravity models for Didymos and Dimorphos, solar radiation pressure with combined occultations, and ephemerides-based third-body perturbations. The ejecta initial conditions were generated using observation-constrained velocity--size distributions and ejecta-cone geometries. In total, \(20\) million independent ejecta trajectories were integrated to statistically characterise the ejecta fate, long-term evolution, and surface accretion.

The simulation results show that escape from the binary system is the dominant outcome for DART-generated ejecta. More than \(93.5\%\) of the particles escape from the system within the two-year propagation time span, whereas only approximately \(0.002\%\) remain in the near-binary environment over extended durations. Although this retained fraction is minute, it may still be dynamically meaningful, particularly for larger and lower-speed fragments that can remain associated with the binary system. Surface accretion is found to be highly non-uniform and strongly dependent on the initial ejecta-cone geometry. On Dimorphos, the deposited layer is estimated to reach the order of \(1.5~\mathrm{mm}\) at mid-to-low latitudes. On Didymos, the accreted layer is mostly thinner than \(0.3~\mathrm{mm}\), but can reach approximately \(3\)--\(11.5~\mathrm{mm}\) within a localised high-density region. The high-density accretion region is particularly associated with cases in which the initial ejecta direction is close to the anti-DART-impact direction.

Overall, the results indicate that most DART-generated ejecta are removed from the binary system, while a small but dynamically meaningful subset can remain near the system or accrete onto the asteroid surfaces. The resulting surface accretion distribution is governed primarily by the initial ejecta-cone geometry, especially the cone-axis direction, while particle size and cone-angle range provide secondary modulation. These findings highlight the importance of accurately constraining the ejecta-cone geometry when modelling the distribution of DART-generated material. The predicted fate statistics and accretion-density patterns may provide useful dynamical context for interpreting future observations by ESA's Hera mission and for assessing the post-impact evolution of the Didymos binary system.

\begin{acknowledgements}
    This work was supported by a UKRI Future Leaders Fellowship [grant number MR/W009498/1] awarded to S. Soldini with the REMORA project (REndezvous Mission for Orbital Reconstruction of Asteroids: A fleet of Self-driven CubeSats for Tracking and Characterising Asteroids), which sponsors X. Fu and N. Stronati. X. Fu and S. Soldini gratefully acknowledge the support from A. Escalante-L\'opez, S. Chesley, and S. Boris for their guidance on ESA's Hera and NASA's DART SPICE kernels. X. Fu also acknowledges the University of Liverpool high-performance computing cluster, Barkla, on which the large-scale simulations in this study were performed, and thanks the Barkla team for their technical assistance. 
\end{acknowledgements}

%

\bibliographystyle{aa} 
\bibliography{reference}{}

\end{document}